\documentclass[fleqn,usenatbib]{mnras}
\usepackage[T1]{fontenc}
\usepackage{ae,aecompl}
\usepackage{graphicx}	
\usepackage{amsmath}	
\usepackage{amssymb}	
\usepackage{booktabs}
\usepackage{hyperref}
\usepackage{bm}
\usepackage[normalem]{ulem}
\usepackage{verbatim}

\newcommand{\cms}{\text{cm~s$^{-1}$}}

\newcommand{\gpercmcu}{\text{g~cm$^{-3}$}}
\newcommand{\msun}{\text{M$_\odot$}}
\newcommand{\msunperyr}{\text{M$_\odot$~yr$^{-1}$}}

\defcitealias{2017MNRAS.468.3850S}{Paper I}
\defcitealias{2018MNRAS.477.1239S}{Paper II}

\title[Ring and gap formation in 3D]{The formation of rings and gaps in wind-launching non-ideal MHD disks: three-dimensional simulations}
\author[Suriano et al.]{
Scott S. Suriano$^{1,2}$\thanks{E-mail: suriano@ea.c.u-tokyo.ac.jp}, Zhi-Yun Li$^{2}$, Ruben Krasnopolsky$^{3}$, Takeru K. Suzuki$^{1}$\newauthor~and Hsien Shang$^{3}$\\
$^{1}$School of Arts \& Sciences, University of Tokyo, 3-8-1 Komaba, Meguro, Tokyo 153-8902, Japan\\
$^{2}$Department of Astronomy, University of Virginia, Charlottesville, VA 22904, USA \\
$^{3}$Academia Sinica, Institute of Astronomy and Astrophysics, Taipei 10617, Taiwan\\
}

\date{Accepted XXX. Received YYY; in original form ZZZ}

\pubyear{2018}

\begin{document}
\label{firstpage}
\pagerange{\pageref{firstpage}--\pageref{lastpage}}
\maketitle

\begin{abstract}
Previous axisymmetric investigations in two dimensions (2D) have shown that rings and gaps develop naturally in non-ideal magnetohydrodynamic (MHD) disk-wind systems, especially in the presence of ambipolar diffusion (AD). Here we extend these 2D simulations to three dimensions (3D) and find that rings and gaps are still formed in the presence of a moderately strong ambipolar diffusion. The rings and gaps form from the same basic mechanism that was identified in the 2D simulations, namely, the redistribution of the poloidal magnetic flux relative to the disk material as a result of the reconnection of a sharply pinched poloidal magnetic field lines. Thus, the less dense gaps are more strongly magnetized with a large poloidal magnetic field compared to the less magnetized (dense) rings. The rings and gaps start out rather smoothly in 3D simulations that have axisymmetric initial conditions. Non-axisymmetric variations arise spontaneously at later times, but they do not grow to such an extent as to disrupt the rings and gaps. These disk substructures persist to the end of the simulations, lasting up to 3000 orbital periods at the inner edge of the simulated disk. The longevity of the perturbed yet still coherent rings make them attractive sites for trapping large grains that would otherwise be lost to rapid radial migration due to gas drag. As the ambipolar diffusivity decreases, both the disk and the wind become increasingly turbulent, driven by the magnetorotational instability, with tightly-wound spiral arms becoming more prominent in the disk.
\end{abstract}

\begin{keywords}
accretion, accretion discs -- magnetohydrodynamics (MHD) -- ISM: jets and outflows -- protoplanetary discs
\end{keywords}

\section{Introduction}

State-of-the-art observational facilities are providing increasingly stringent constraints on the physical properties of circumstellar disks. Specifically, the Atacama Large Millimeter/submillimeter Array (ALMA) has shown that a large number of circumstellar disks have detailed radial and azimuthal substructures \citep{2015ApJ...808L...3A,2016ApJ...820L..40A,2016ApJ...818L..16Z,2016ApJ...819L...7N,2016Sci...353.1519P,2016PhRvL.117y1101I,2016Natur.535..258C,2017A&A...597A..32V,2017A&A...600A..72F,2018A&A...610A..24F,2018MNRAS.475.5296D}. How the various observed structures form remains undetermined, though a number of promising physical mechanisms have been proposed, including planet-disk interactions \citep{2015ApJ...809...93D,2017ApJ...843..127D,2017ApJ...850..201B}, rapid pebble growth at the condensation fronts of abundant volatile species \citep{2015ApJ...806L...7Z}, the pileup of volatile ices in sintering zones just outside snow lines \citep{2016ApJ...821...82O}, sharp changes in the disk viscosity at the boundaries of non-turbulent `dead zones' \citep{2015A&A...574A..68F,2016A&A...590A..17R}, magnetic self-organization through zonal flows \citep{2016A&A...589A..87B,2017A&A...600A..75B,2018A&A...617A.117R,2018ApJ...865..105K}, variable magnetic disk winds \citep{2017MNRAS.468.3850S,2018MNRAS.477.1239S}, outward dust migration forced by preferential gas removal from the inner disk through a disk wind \citep{2018ApJ...865..102T}, and the secular gravitational instability \citep{2014ApJ...794...55T,2016AJ....152..184T}. The disk substructures (i.e., rings and gaps, spirals, vortices) have an undoubtedly important influence on the concentration and growth of dust grains in the disks around young stars. They could, for example, prevent the fast radial migration of large grains through disks by trapping grains in the pressure maxima they create, possibly even early on in the disk lifetime.

There is growing evidence that large-scale magnetic fields thread circumstellar disks. These magnetic fields could be responsible for launching disk winds (e.g., \citealt{2016Natur.540..406B,2018arXiv181106544B}) and driving accretion via a laminar magnetic wind torque (e.g., \citealt{2013ApJ...769...76B,2013ApJ...775...73S}). The motivation to explore how magnetic fields can create disk substructures is clear, though the idea that magnetic flux can concentrate in localized regions of disks is not new. For example, magnetic self-organization is seen in disks the form of `zonal flows', or radially banded regions of sub- and super-Keplerian rotation. These flows are generated by large-scale variations in the magnetic stresses in MRI turbulent disks, and the resulting radial variations in density and magnetic flux are long-lived as the pressure gradient is balanced by the Coriolis force \citep{2009ApJ...697.1269J}. Subsequent numerical studies have shown that zonal flows can be driven by the anisotropic nature of the non-ideal Hall \citep{2013MNRAS.434.2295K,2015ApJ...798...84B,2016A&A...589A..87B} and ambipolar diffusivities \citep{2014ApJ...784...15S,2014ApJ...796...31B,2017A&A...600A..75B}. Similarly, zonal flows can develop at the edge of an MRI `dead zone' in response to the creation of a density/pressure maximum inside the slowly accreting dead zone \citep{2010A&A...515A..70D,2015A&A...574A..68F,2016A&A...590A..17R}. These regions of the disk can also be unstable to the Rossby wave instability, forming vortices which last for tens of local orbital periods. Finally, the formation of radial density structures due to a strong poloidal magnetic field that drives a magnetocentrifugal wind \citep{1982MNRAS.199..883B} was demonstrated by \citet{2012A&A...548A..76M} in local 2D (shearing-box) simulations.

So far, the formation of radial substructures via MHD disk winds have only been explored in two dimensions (2D), which necessitates that the structures are axisymmetric (e.g., \citealt{2017MNRAS.468.3850S,2018MNRAS.477.1239S}, herein \citetalias{2017MNRAS.468.3850S,2018MNRAS.477.1239S}, respectively). In the 2D simulations of \citetalias{2018MNRAS.477.1239S}, the formation of rings and gaps on observable scales ($r\sim10$~au) proceeds due to the effects of ambipolar diffusion (AD), where AD is the most important non-ideal MHD effect. The rings and gaps are naturally produced in the presence of a significant poloidal magnetic field and a relatively strong ambipolar diffusion, from which a relatively laminar disk-wind system develops (see also earlier work by \citealt{1993ApJ...410..218W,1995A&A...295..807F,1996ApJ...465..855L}). The mechanism is driven by reconnection of the highly pinched poloidal magnetic field in a thin midplane current sheet where the reconnection leads to the weakening of the poloidal field in some regions, which accrete more slowly and form rings, and field concentration in others, which accrete efficiently and open gaps.

This work explores the formation of rings and gaps in circumstellar disks by magnetic disk winds in the presence of ambipolar diffusion in three dimensions (3D), the logical next step to determine whether and if so, how substructures develop in magnetically coupled disk-wind systems. We find that prominent rings and gaps are still formed in 3D, apparently through the same mechanism of a thin midplane current sheet leading to reconnection and, therefore, the redistribution of the poloidal magnetic flux relative to disk matter. However, the reconnection of the field need not be restricted to the meridian ($r\theta$-) plane. Unlike the 2D (axisymmetric) case, the reconnection of the toroidal ($\phi$) component of the magnetic field is allowed, and can in principle lead to interesting non-axisymmetric structures.

The rest of the paper is organized as follows. In Section~\ref{sec:setup}, we describe the simulation setup, including the initial and boundary conditions. The results of the simulations are presented in Sections~\ref{sec:results} and \ref{sec:param}. Section~\ref{sec:results} focuses on a high-resolution reference simulation where the AD Elsasser number at the inner edge of the disk is $\Lambda_0=0.25$ (see Section~\ref{sec:AD} for a description of this quantity). Other lower resolution simulations with different AD Elsasser numbers and magnetic field strengths are discussed in Section~\ref{sec:param}. We discuss the implications of this work in Section~\ref{sec:discuss} and summarize the main results in Section~\ref{sec:conc}.

\section{Simulation setup}\label{sec:setup}

The simulation setup is nearly identical to \citetalias{2018MNRAS.477.1239S}, except that the assumption of axisymmetry is now removed. We will summarize the main features of the setup here, and refer the reader to \citetalias{2018MNRAS.477.1239S} for details. We use the \textsc{ZeusTW} code \citep{2010ApJ...716.1541K} to solve the time-dependent non-ideal magnetohydrodynamic (MHD) equations in spherical polar coordinates ($r,\theta,\phi$) including ambipolar diffusion (see equations (1)-(4) of \citetalias{2018MNRAS.477.1239S}, with the Ohmic resistivity, $\eta_O$, set to zero).

\subsection{Initial conditions: disk, corona, and magnetic field}

We start with a cold, geometrically thin, magnetized disk, surrounded by a hotter corona that is in hydrostatic equilibrium with the disk surface. We characterize the disk by the dimensionless parameter $\epsilon = h/r = c_s/v_K =0.05$, where $h$ is the disk scale height, $c_s$ the isothermal sound speed, and $v_K$ is the Keplerian speed, which is fixed by the stellar mass (taken to be 1~\msun). We set the initial disk (half) opening angle to $\theta_0=\arctan(2\epsilon)=5.7^\circ$. Within the disk defined by (twice) this opening angle, we prescribe a density distribution
\begin{equation}
\rho_d(r,\theta,\phi) = \rho_{0} \left(\frac{r}{r_0}\right)^{-3/2} \exp \left(-\frac{\cos^2\theta}{2 \epsilon^2}\right),
\end{equation}
where the subscript `0' refers to values on the disk midplane at the inner radial boundary $r_0=1$~au. We choose a fiducial density scaling of $\rho_0=1.3\times10^{-10}~\gpercmcu$, corresponding to a relatively large initial disk mass of about 0.1~M$_\odot$ for the 100 au disk. It is possible to rescale the simulation to other choices of $\rho_0$, keeping dimensionless quantities such as the plasma-$\beta$ fixed. For example, if we reduce the density scaling $\rho_0$ by a factor of 10 (to $1.3\times 10^{-11}~\gpercmcu$), both the disk mass and mass accretion rate would be reduced correspondingly by the same factor of 10 and the magnetic field strength by a factor of $\sqrt{10}$ if the stellar mass is kept unchanged. The density distribution of the corona is set by pressure balance across the disk surface (see equation~(9) of \citetalias{2018MNRAS.477.1239S}). As before, we choose an adiabatic index close to unity ($\gamma=1.01$) so that the temperature of a given fluid parcel stays nearly constant as it moves around in the simulation domain.

The initially axisymmetric magnetic field is computed from the magnetic flux function $\Psi$ using the following equations: 
\begin{equation}
B_r=\frac{1}{r^2\sin{\theta}}\frac{\partial\Psi}{\partial\theta},\label{eq:br}
\end{equation}
\begin{equation}
B_\theta=-\frac{1}{r\sin\theta}\frac{\partial\Psi}{\partial r}.\label{eq:btheta}
\end{equation}
Following \citet{2007A&A...469..811Z}, we adopt 
\begin{equation}
\Psi(r,\theta) = \frac{4}{3}r_0^2 B_{\mathrm{p},0}\left(\frac{r\sin\theta}{r_0}\right)^{3/4} \frac{0.5^{5/4}}{\left(0.5^{2}+\cot^2\theta\right)^{5/8}}\label{eq:psi},
\end{equation}
which describes a large-scale, ordered, poloidal magnetic field that threads the disk midplane vertically and bends gradually outward away from the midplane \citep{2014ApJ...793...31S}. The field strength scaling, $B_{\mathrm{p},0}$, is set by specifying the initial plasma-$\beta$ (the ratio of the thermal to magnetic pressure) on the disk midplane, which is $10^3$ for all but one simulation.

\subsection{Ambipolar diffusion}\label{sec:AD}

The coefficient for ambipolar diffusion can in principle be computed through detailed chemical networks (e.g., \citealt{2009ApJ...701..737B}). As a first step toward a comprehensive model, we will simply parametrize the density of ions as
\begin{equation}
\rho_i = \rho_{i,0} f(\theta) \left(\frac{\rho}{\rho_0}\right)^{\alpha_{AD}},
\end{equation}
where
\begin{equation}\label{eq:ftheta}
  f(\theta) =
  \begin{cases}
    \exp\left(\frac{\cos^2(\theta+\theta_0)}{2 \epsilon^2}\right) & \theta<\pi/2-\theta_0 \\
    1                                                             & \pi/2-\theta_0\leq\theta\leq\pi/2+\theta_0 \\
    \exp\left(\frac{\cos^2(\theta-\theta_0)}{2 \epsilon^2}\right) & \theta>\pi/2+\theta_0.
  \end{cases}
\end{equation}
The function $f(\theta)$ is chosen to mimic the increase in ionization by high energy photons (UV and X-rays) from the central young star in addition to cosmic rays near the disk surface (e.g., \citealt{1981PASJ...33..617U,2011ApJ...735....8P,2017MNRAS.472.2447G}). For simplicity, we adopt  $\alpha_{AD}=0.5$, the value expected in the case where the volumetric cosmic ray ionization rate is balanced by the recombination rate of ions and electrons, under the constraint of charge neutrality (i.e., $\zeta n\propto n_e n_i \propto n_i^2$, where $\zeta$ is the cosmic ray ionization rate per hydrogen nucleus; see page 362 of \citealt{1992phas.book.....S}). An obvious future improvement of this work is to include a more self-consistent treatment of the ionization, perhaps along the lines of, e.g., \citet{2018arXiv181012330W}.

The magnitude of ambipolar diffusion is often characterized by the dimensionless ambipolar Elsasser number,
\begin{equation}
\Lambda=\frac{\gamma_i \rho_i}{\Omega},
\end{equation}
where $\gamma_i $ is the frictional drag coefficient between ions and neutrals. Physically, the Elsasser number is the collision frequency of a neutral particle in a sea of ions of density $\rho_i$, normalized to the Keplerian orbital frequency. Therefore, the larger the Elsasser number is, the better coupled the bulk neutral disk material is to the ions (and the magnetic field attached to them). For our reference simulation (Section~\ref{sec:results}), we choose $\Lambda_0=0.25$ at the inner boundary on the disk midplane, but we will vary this parameter to gauge its effects on the coupled disk-wind system (Section~\ref{sec:param}). Note that with our choice of $\alpha_{AD}=0.5$, the Elsasser number is proportional to radius as $r^{3/4}$, i.e., the disk material at larger radii is better coupled to the magnetic field than at smaller radii in the simulations. Specifically, we have $\Lambda=1.4$ at 10~au and $7.9$ at 100~au for the reference simulation. For the simple scaling of $\rho_i=C\rho^{1/2}$ given in \citet{1992phas.book.....S} where $C=3\times 10^{-16} \left(\zeta/10^{-17}\mathrm {s}^{-1}\right)^{1/2}~\mathrm{g}^{1/2}~\mathrm{cm}^{-3/2}$ and $\gamma_i =3.5\times 10^{13}~\mathrm{cm}^3~\mathrm{g}^{-1}~\mathrm{s}^{-1}$, the Elsasser number in the reference simulation corresponds to a cosmic ray ionization rate of $\zeta=1.8\times 10^{-18}~\mathrm{s}^{-1}$ for the fiducial disk mass of 0.1~\msun. For a less massive disk of 0.01~\msun, the cosmic ray ionization rate would need to be increased by a factor of 10, to $\zeta=1.8\times 10^{-17}~\mathrm{s}^{-1}$, to keep the dimensionless Elsasser number unchanged.

\subsection{Grid}\label{sec:grid}

The equations are solved for $r\in{[1,100]}$~au, $\theta\in{[0,\pi]}$, and $\phi\in{[0,2\pi]}$ with a resolution of $n_r\times n_\theta\times n_\phi= 300\times270\times270$ for a high-resolution reference run. A `ratioed' grid is used in the radial direction such that $dr_{i+1}/dr_i$ is constant and $r_{i+1}=r_i+dr_i$. The grid spacing at the inner edge is set as $dr_0=2.0~r_0d\theta_\mathrm{mid}$. The $\theta$ grid is separated into three $60^\circ$ blocks, the middle of which, from $\theta=60^\circ$ to $120^\circ$, is uniform with 180 cells for a resolution of $0.33^\circ$ per cell or 17 cells from the disk midplane to the initial disk surface at two scale heights. The first and last $\theta$ grid blocks are ratioed grids with the cell size matched to the resolution of the middle block at their boundary and increasing towards the poles where the cells reach a maximum width of $3.4^\circ$. The $\phi$ grid is uniform with 270 cells. This results in the cells at the inner boundary on the midplane being a box with dimensions of 2:1:4 in the $r:\theta:\phi$ directions. For a parameter survey, we perform several additional simulations for a range of AD coefficients and a weaker magnetic field strength but with a lower resolution of $n_r\times n_\theta\times n_\phi= 200\times180\times180$ to lower the computational costs.

\subsection{Boundary conditions}\label{sec:bc}

Both the inner and outer radial boundaries use the standard outflow condition, where scalar quantities and the $r$ and $\phi$ components of vector quantities are copied from the last active grid cell into ghost zones. The radial components of the vector quantities that are directed towards the active simulation domain are set to zero in the ghost zones. However, $B_\phi$ is set to zero on the inner radial boundary, as it is taken to be non-rotating. The reflection boundary condition is used on the polar axis and the $\theta$ grid is shifted slightly off of $\theta=0$ to avoid the coordinate singularity. The $\phi$ component of the magnetic field is also set to zero on the polar axis.

In the polar regions where the field lines have footpoints on the non-rotating inner boundary rather than threading the rotating disk, matter can easily stream along the magnetic fields lines towards the inner radial boundary due to the gravity of the central star. This evacuates the polar region leading to such a large Alfv\'en speed that the simulation time step becomes prohibitively small. We limit the Alfv\'en speed in the problematic cells by adding mass, which sometimes creates high densities in the polar region (see Fig.~\ref{fig:global}). Most of the added mass falls quickly into the central hole and does not strongly affect the disk away from the polar region.

\begin{table}
  \centering
  \caption{Model Parameters}
  \label{tab:sims3D}
  \begin{tabular}{l c c c c c c c}
	\hline
    \hfill & $\beta$ & $\Lambda_0$ & Resolution \\
    \hline
    \textbf{reference}  & $10^3$ & 0.25  & $300\times270\times270$ \\
    ad-els0.05 & $10^3$ & 0.05  & $200\times180\times180$  \\
    ad-els0.25 (low-res reference) & $10^3$ & 0.25  & $200\times180\times180$  \\
    ad-els1.25 & $10^3$  & 1.25 & $200\times180\times180$  \\
    ad-els$\infty$ (ideal) & $10^3$  & $\infty$ & $200\times180\times180$  \\
    beta1e4    & $10^4$ & 0.25 & $200\times180\times180$ \\
   \hline
  \end{tabular}
\end{table}

\section{Reference Simulation}\label{sec:results}

We run a small suite of 3D simulations to examine the formation of substructure in disks that are threaded by a large-scale, ordered poloidal magnetic field. The simulations differ in either the Elsasser number (i.e., the magnetic coupling), the magnetic field strength, or the grid resolution (see Table~\ref{tab:sims3D} for a list of models). This section first focuses on the high-resolution simulation (termed the `reference' simulation hereafter), which has an Elsasser number at the inner edge of the disk of $\Lambda_0=0.25$ and a midplane plasma-$\beta$ of $10^3$. Other lower resolution simulations with a range of Elsasser number and a different value of the magnetic field strength will be discussed in the next section.

\subsection{The coupled disk-wind system: a global view}\label{sec:global}

\begin{figure*}
\centering
\includegraphics[width=2.0\columnwidth]{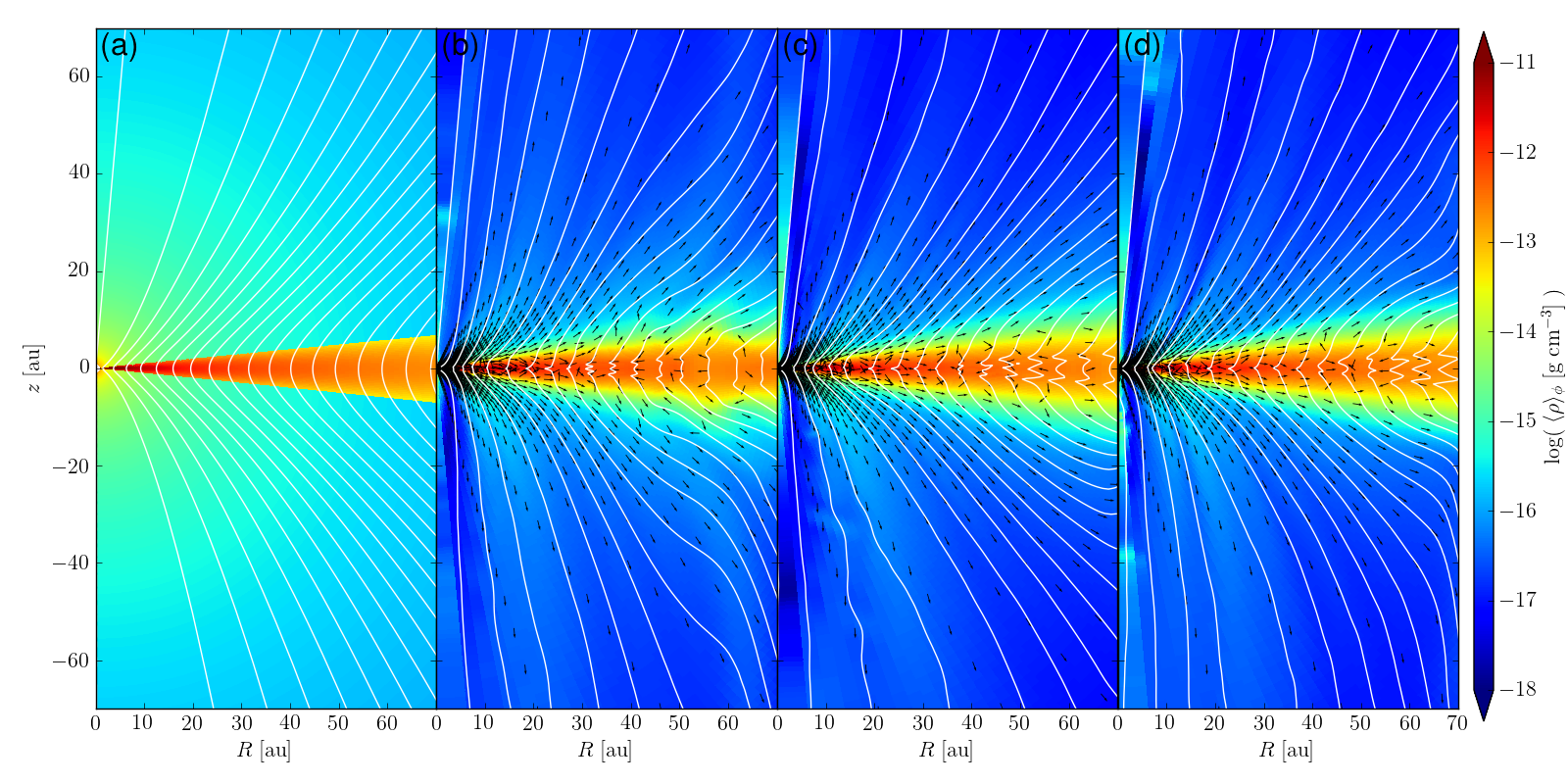}
\caption{The azimuthally averaged density (colour contours) in the 3D reference simulation. Also shown are the `effective' poloidal magnetic field lines (white) and the poloidal velocity unit vectors (black). The panels (a)-(d) from left to right correspond to simulation times of 0, 500, 1000, and 1500 inner orbital periods. (See the supplementary material in the online journal for an animated version of this figure.)}
\label{fig:global}
\end{figure*}

In order to make a connection between these 3D simulations and the 2D (axisymmetric) simulations presented in \citetalias{2018MNRAS.477.1239S} (where the simulation results have relatively intuitive physical interpretations), we start our discussion with azimuthally averaged quantities, especially the poloidal components of the magnetic field, $B_r$ and $B_\theta$. A great advantage of the axisymmetric simulations is that these two field components can be expressed in terms of a magnetic flux function $\Psi$ (through equations~\ref{eq:br} and \ref{eq:btheta}). This makes it easy to define poloidal magnetic lines -- lines of constant magnetic flux function $\Psi$. The poloidal field lines play an important role in our analysis of the 2D (axisymmetric) simulations, particularly in illustrating the radial pinching and reconnection of the magnetic field lines (e.g., see Fig.~1, 7, and 8 of \citetalias{2018MNRAS.477.1239S}). For 3D simulations that are generally non-axisymmetric, the poloidal field components $B_r$ and $B_\theta$ at any given point can no longer be expressed in terms of a scalar function. As a result, analyzing the behavior of the magnetic field becomes much more challenging. Nevertheless, if we define two azimuthally averaged poloidal field components, $\overline{B}_r$ and $\overline{B}_\theta$, it is easy to show that they can again be expressed in terms of a scalar function $\overline{\Psi}$ through equations~(\ref{eq:br}) and (\ref{eq:btheta}), respectively, just as in the axisymmetric case. Physically, $2\pi\overline{\Psi}$ at any given point is the total magnetic flux enclosed within a circle centered on the axis that passes through that point; it has the same physical meaning as $2\pi\Psi$ in the axisymmetric case. We will take advantage of this property of the azimuthally averaged poloidal field components in the spherical polar coordinate system adopted in this work, and plot together with other azimuthally averaged quantities contours of constant $\overline{\Psi}$, which are azimuthally averaged (or `effective') poloidal magnetic field lines.

Figure~\ref{fig:global} gives a global perspective of the reference simulation. This figure shows azimuthally averaged snapshots of the density (colour map), effective poloidal magnetic field lines (white), and poloidal velocity unit vectors (black) out to $R=70$~au at four times ($t/t_0=0,500,1000,1500$, where $t_0=1$~yr is the orbital period at the inner edge of the disk). It shows that a wind is launched over most of the disk surface, except very close to the polar axis. Upon first inspection, the magnetic field in the disk has two distinct modes of evolution separated at about $r\sim30$~au in the second frame shown in Fig.~\ref{fig:global} (at $t/t_0=500$). Within this radius, the magnetic field is dragged radially inward near the midplane leading to a sharp radial pinch. This is caused  by the vertical steepening of a midplane current sheet due to AD as described in detail in \citetalias{2018MNRAS.477.1239S} (see their Section~3.3.1). The current peaks where the toroidal magnetic field changes sign from positive below the disk to negative above the disk and the effects of AD further steepen the magnetic gradient near the magnetic null \citep{1994ApJ...427L..91B}. This is the same phenomenon that leads to the magnetic reconnection of the poloidal magnetic field in 2D. Outside of $r\sim30$~au, `channel-flow like' structures are evident. This is because the disk material is better coupled to the magnetic field at larger radii, since the radial dependence of the Elsasser number goes as $\Lambda\propto r^{3/4}$. The demarcation radius between these two types of disk accretion moves outward with time, because it takes time for the current layers created at the disk surfaces to migrate towards and converge at the disk midplane as the induced toroidal magnetic pressure gradient from the winding of the initially poloidal field grows (see Section~\ref{sec:ringgap}). By the last frame shown in Fig.~\ref{fig:global} at $t/t_0=1500$, the magnetic field lines are radially pinched by the midplane accretion layer out to approximately 40~au.

\begin{figure*}
\centering
\includegraphics[width=2.0\columnwidth]{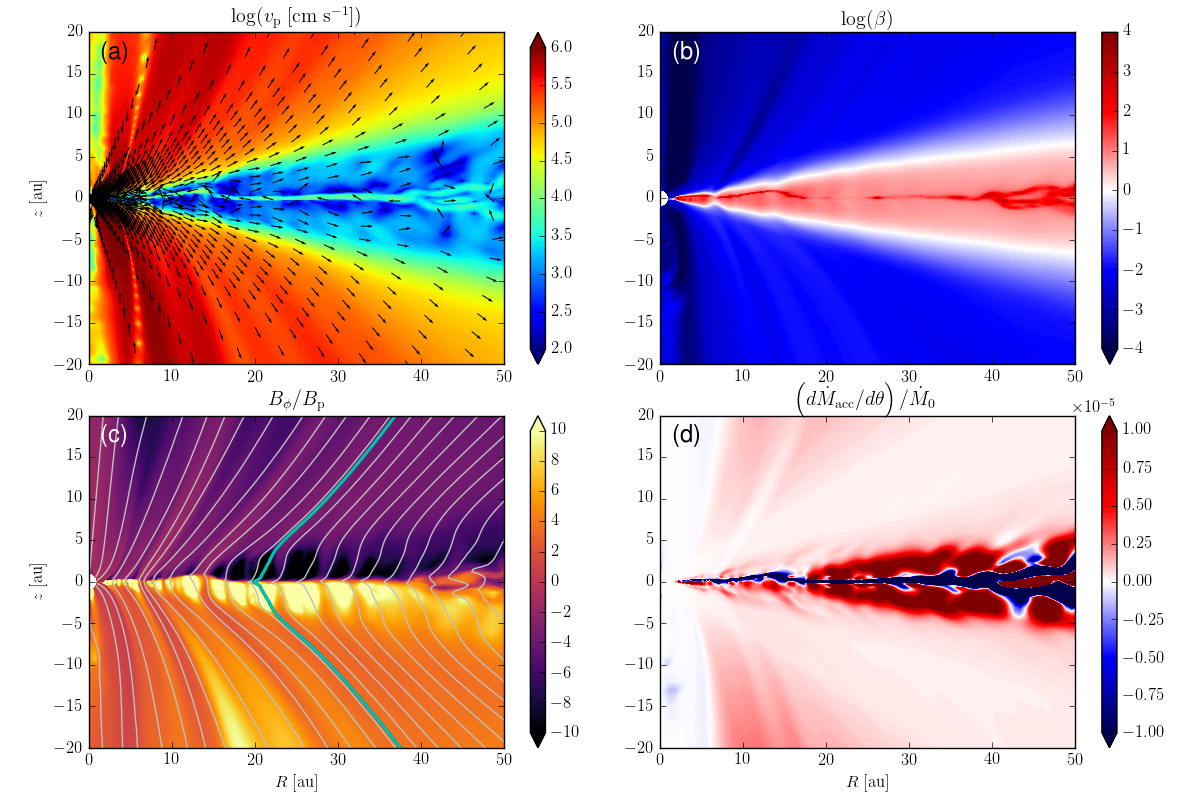}
\caption{The reference simulation at a representative time $t/t_0=1500$. The panels show the following $\phi$-averaged quantities: (a) the logarithm of the poloidal velocity (\cms) with poloidal velocity unit vectors; (b) the logarithm of plasma-$\beta$; (c) the ratio of the toroidal to the poloidal magnetic field strength with the effective poloidal field lines (gray lines; the cyan line highlights a representative effective field line with a midplane footpoint at a radius of $20$~au); (d) the differential mass accretion rate integrated over $\phi$, i.e., $d\dot{M}_\mathrm{acc}/d\theta=\int_0^{2\pi}\rho v_r r^2 \sin{\theta} d\phi$, normalized to $\dot{M}_0=r_0^2\rho_0 c_{s,0}$.}
\label{fig:panels}
\end{figure*}

To further illustrate the overall structure of the disk and its connection to the wind, Fig.~\ref{fig:panels} shows several $\phi$-averaged quantities of the disk-wind system. Panel (a) shows the poloidal velocity where the fastest accretion through the disk is limited to a thin current layer near the midplane with accretion speeds on the order of 10~$\mathrm{m~s^{-1}}$. Above and below this accretion layer, the poloidal velocity is directed outward and the velocity is approximately one order of magnitude less than in the accretion layer. Also in the thin accretion layer, the value of plasma-$\beta$ peaks at values approaching $10^4$ (Fig.~\ref{fig:panels}b). Similarly, the mass accretion rate is largest in this thin layer as shown in Fig.~\ref{fig:panels}(d), which plots the $\phi$-integrated radial mass accretion rate per unit polar angle, $d\dot{M}_\mathrm{acc}/d\theta=\int_0^{2\pi}\rho v_r r^2 \sin{\theta} d\phi$.  Again, it is clear that the disk mass moves outward above and below the midplane accretion layer and further extends to a tenuous disk wind beyond the disk surfaces. The fast accretion takes place through a strong current layer where the $\phi$ component of the magnetic field changes sign.\footnote{Note that the midplane current layer is not symmetric about $\theta=\pi/2$ for radii less than approximately 18~au (see Fig.~\ref{fig:panels}d and \citealt{2017A&A...600A..75B,2017ApJ...845...75B}).} The field reversal leads to a low value of $B_\phi$ in the accretion layer, which is the reason why its plasma-$\beta$ is high. It can be seen more clearly in Fig.~\ref{fig:panels}(c), which plots the ratio of the toroidal to the poloidal magnetic field. We see that the magnetic field is pinched radially inward where the toroidal magnetic field changes sign from positive below the disk to negative above the disk. Also, we see a variation in the ratio $B_\phi/B_\mathrm{p}$ (i.e., the degree of magnetic field twisting) as a function of radius, where the poloidal magnetic field lines concentrate in some regions (where the ratio is relatively low) while the toroidal magnetic field is more dominant in others (where the ratio is higher). The cause of this alternating pattern of different degrees of field twisting will be discussed in the next section.

\begin{figure*}
\centering
\includegraphics[width=1.0\columnwidth]{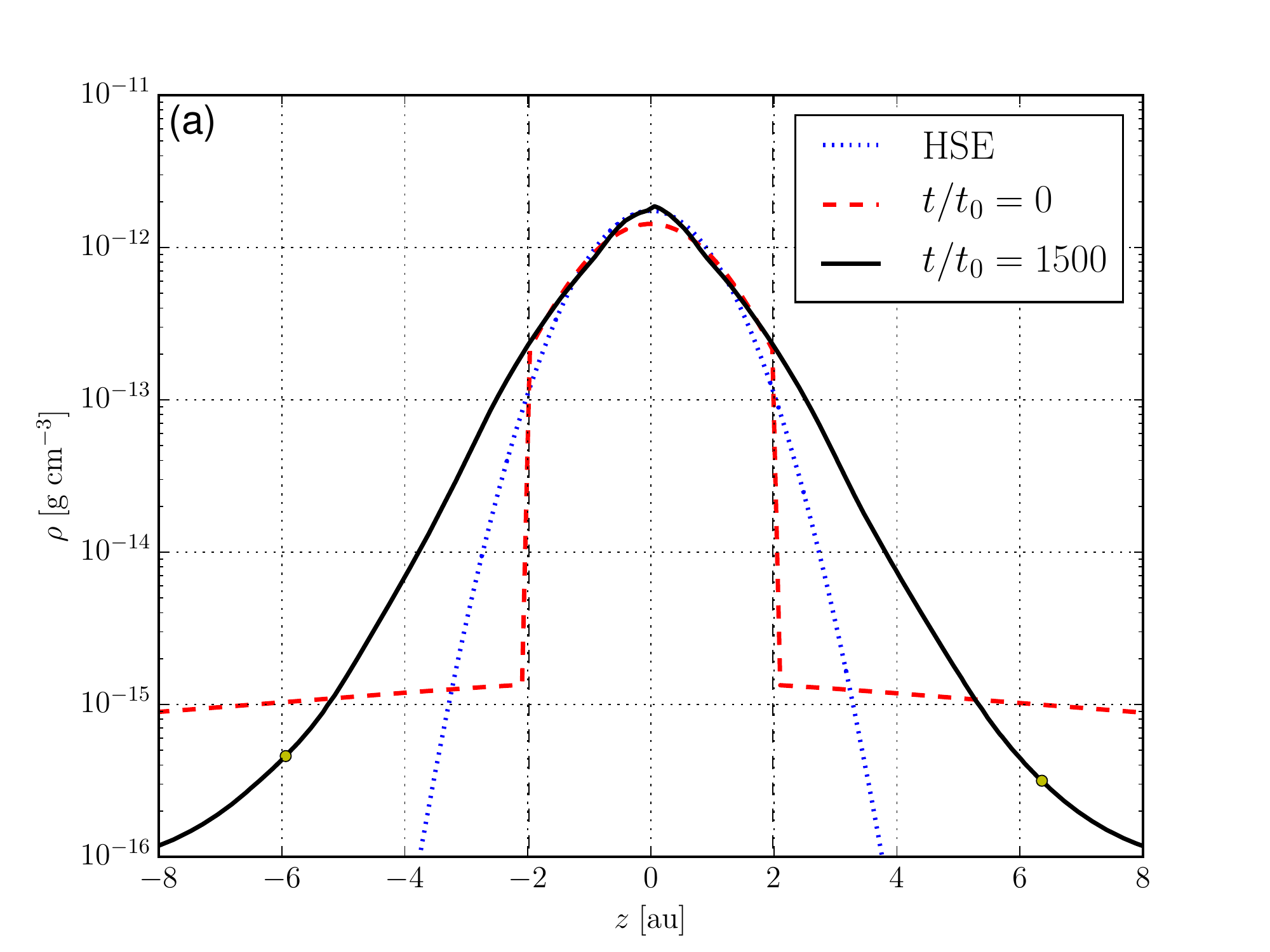}
\includegraphics[width=1.0\columnwidth]{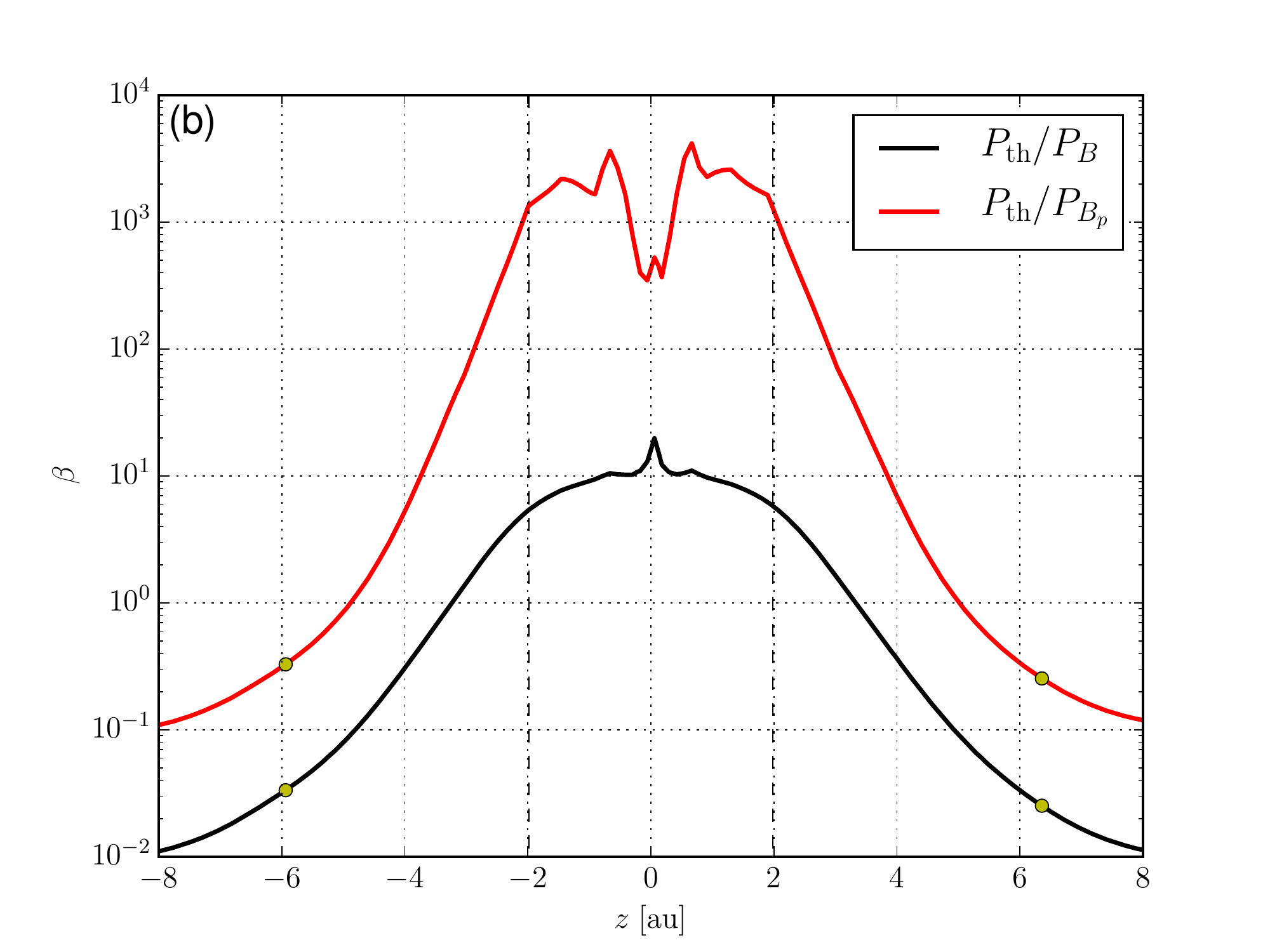}
\includegraphics[width=1.0\columnwidth]{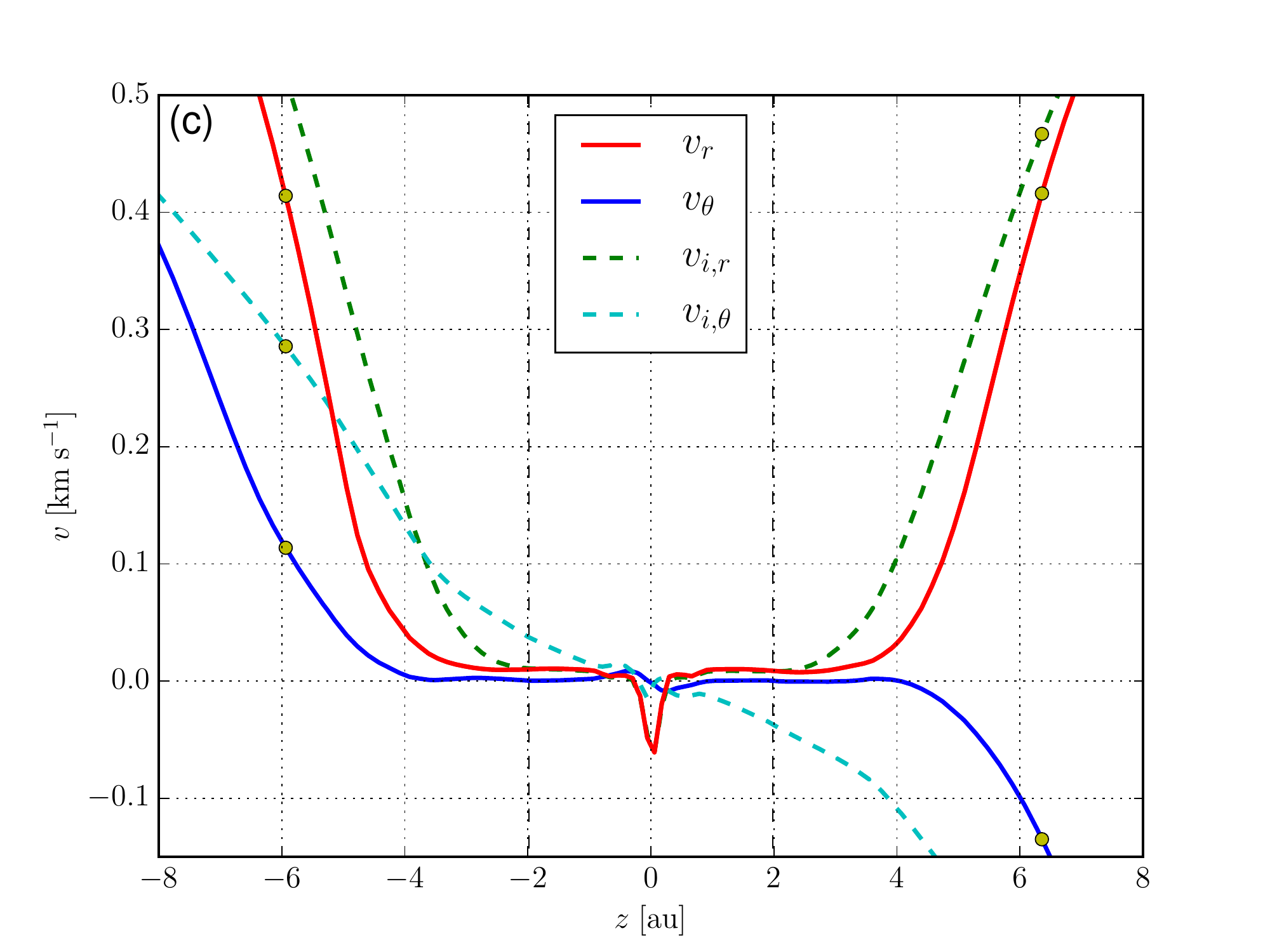}
\includegraphics[width=1.0\columnwidth]{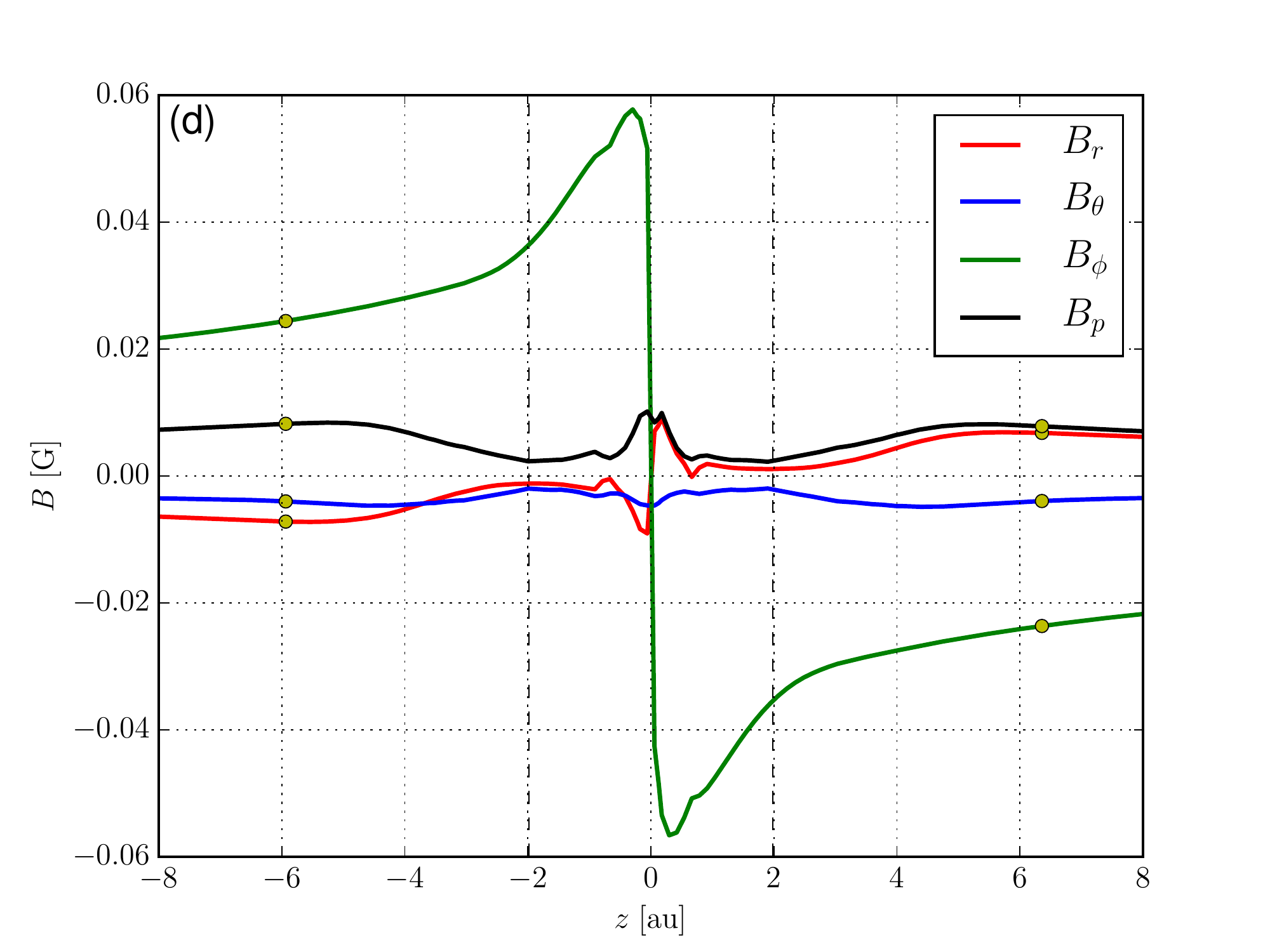}
\caption{Azimuthally averaged quantities along a $\phi$-averaged poloidal magnetic field line plotted as a function of $z$ at $t/t_0=1500$. The magnetic field line passes through the disk midplane at a radius of $r=20$~au and it is highlighted in Fig.~\ref{fig:panels}(c) in cyan. The panels plot: (a) the density for $t/t_0=0$ (red dashed line), 1500 (solid black line), and the hydrostatic equilibrium distribution given the midplane density at $t/t_0=1500$ (blue dotted line); (b) plasma-$\beta$ for the total (black) and poloidal (red) magnetic field strength; (c) the poloidal components of the neutral (solid lines) and ion velocities (dashed lines); (d) the magnetic field components. The sonic point, where the poloidal velocity is equal to the adiabatic sound speed, is marked by yellow circles and the initial disk height, $z=\pm2h_0$, is shown by the vertical dashed black lines.}
\label{fig:alongBfield}
\end{figure*}

Figure~\ref{fig:alongBfield} illustrates the properties of the disk and the disk-wind transition region more quantitatively, along a representative effective poloidal field line with a midplane footpoint radius of $20$~au at time $t/t_0=1500$ (shown as the cyan line in Fig.~\ref{fig:panels}c). Panel (a) of Fig.~\ref{fig:alongBfield} shows the distribution of the mass density along the field line, showing that the initially large (two orders of magnitude) drop in density between the initial disk and corona is completely erased by the wind, which connects smoothly to the disk. The two filled circles mark the locations of the sonic points along the field line above and below the disk. Panel (b) shows the distributions of the ratios of the thermal energy density and the energy densities of the total magnetic field and the poloidal field component respectively. It is clear that the winding of the initial poloidal field has increased the magnetic energy in the disk considerably (by two orders of magnitude) over most of the disk, although it remains well below the thermal energy within the boundary of the initial disk (marked by the vertical dot-dashed lines in the figure). The magnetic energy becomes comparable to the thermal energy at $\sim 3$ scale heights, and completely dominates the thermal energy beyond the sonic points (by a factor of 30 or more). Panel (c) shows that the flow acceleration to produce the magnetically-dominated supersonic wind starts to take off above $\sim 4$ scale heights (see the red curve in the panel). Below this height, the material in the disk and the base of the wind expands slowly at a highly subsonic speed, except in a thin layer close to the midplane, where it accretes at a much higher speed. The simultaneous existence of accretion and expansion is one of the key features of the reference coupled disk-wind system. This is expected since the angular momentum extracted by the magnetic torque from the midplane accreting layer is given to the layers above and below it, causing them to expand subsonically initially and eventually as a supersonic wind. Note that the ions have significantly different velocities than neutrals because of ambipolar diffusion (as in the 2D simulations of \citetalias{2018MNRAS.477.1239S}). In particular, their radial velocity is larger than that of neutrals, because the radial component of the magnetic force is directing outward, as required for the wind acceleration. Similarly, in the $\theta$-direction, the ions in the wind are moving faster away from the disk plane than the neutrals, again in response to the wind-driving magnetic force in that direction. The driving force comes from the gradient of the magnetic pressure that is dominated by the toroidal field component $\vert B_\phi\vert$, which decreases away from the disk midplane everywhere, except near the midplane, where it reverses polarity sharply. This sharp reversal of $B_\phi$ is the engine that drives the fast midplane accretion and the resultant pinching of the poloidal magnetic field (see the red line of panel d for the sharp change in the sign of $B_r$ across the midplane).

\begin{figure}
\centering
\includegraphics[width=1.0\columnwidth]{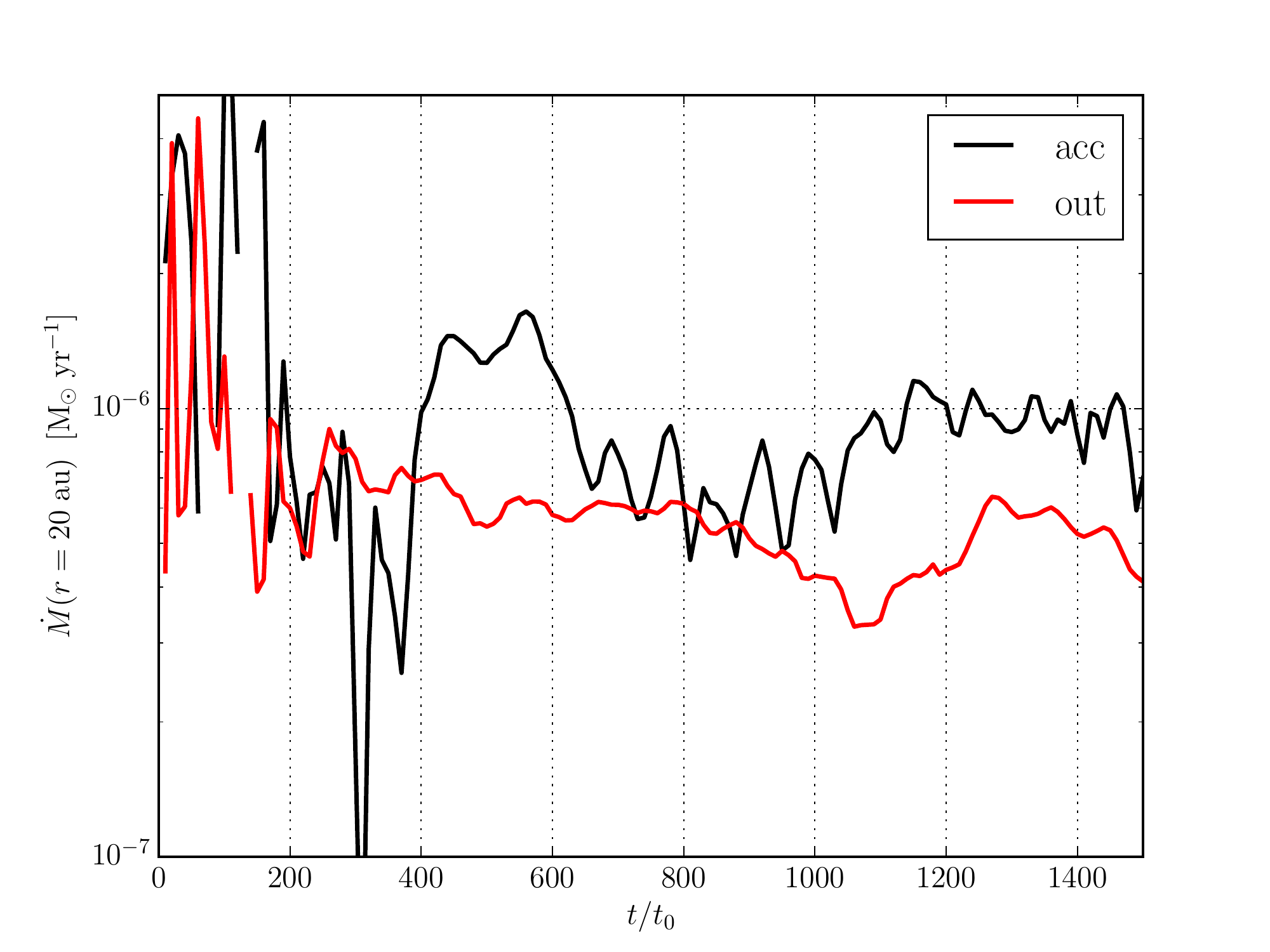}
\caption{The mass accretion and outflow rates (\msunperyr) as a function of time in the 3D reference simulation through a sphere of radius $r=20$~au. The mass accretion rate through the disk ($\vert \pi/2-\theta \vert<2\epsilon$) is shown in black and the total mass outflow rate both above and below the disk ($\vert \pi/2-\theta \vert>2\epsilon$) is shown in red.}
\label{fig:mdot}
\end{figure}

\begin{figure}
\centering
\includegraphics[width=1.0\columnwidth]{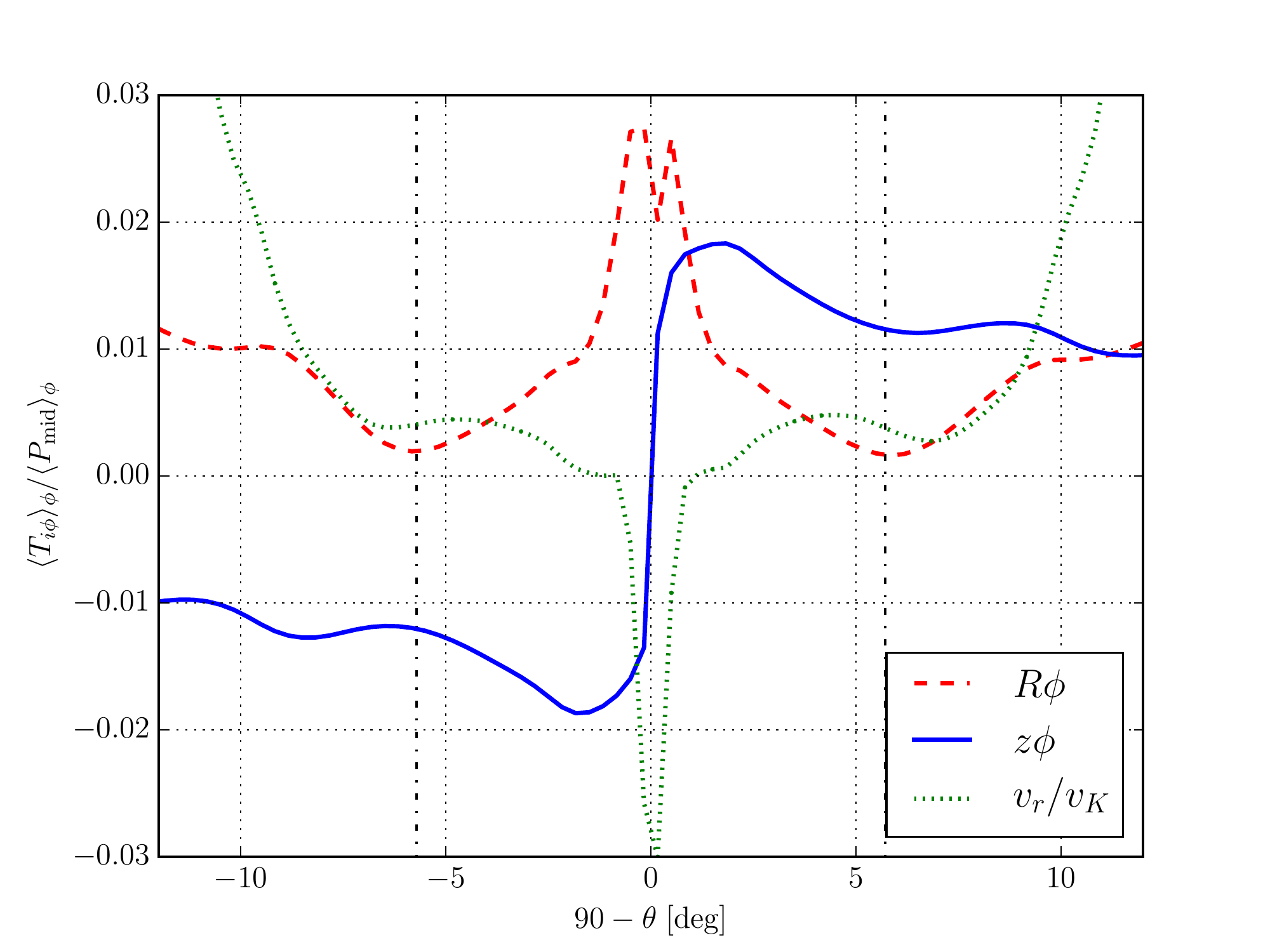}
\caption{The vertical profile of the $\phi$-averaged Maxwell stress tensor normalized to the midplane gas pressure at $r=20$~au for $t/t_0=1500$. The $R\phi$ and $z\phi$ components are shown by the red dashed and blue solid lines, respectively. The $\phi$-averaged radial velocity normalized to the Keplerian velocity at $r=20$~au is plotted for reference (green dotted line). The vertical dot-dashed lines show the initial disk angle of $\theta=\pi/2\pm\theta_0$ (at two scale heights).}
\label{fig:stress}
\end{figure}

Even though the bulk of the disk material is slowly expanding (as shown in Fig.~\ref{fig:panels}a and d, and  Fig.~\ref{fig:alongBfield}c), the disk as a whole is accreting rapidly. This is illustrated in  Fig.~\ref{fig:mdot}, which shows both the mass accretion through the disk at a representative radius $r=20$~au ($\vert\pi/2-\theta\vert<\theta_0$; black line) and the mass outflow rate through a sphere of the same radius excluding the disk region (red line). The mass accretion rate fluctuates with time and settles to a value of $\sim 10^{-6}~\msunperyr$ at later times (for the fiducial choice of the disk mass of 0.1~M$_\odot$). The mass outflow rate approaches approximately $5\times10^{-7}~\msunperyr$. Therefore, more mass is accreted inward through the disk than is ejected away in the wind at this radius. As discussed earlier, the mass accretion and mass loss rates should scale linearly with the adopted disk mass.

Figure~\ref{fig:stress} shows the vertical distribution of the azimuthally averaged Maxwell stress tensor, $T_{i\phi}=-B_iB_\phi/4\pi$ for $i=R,z$ normalized to the midplane gas pressure (similar to the $\alpha$ parameter of \citealt{1973A&A....24..337S}). The $R\phi$ component of the stress tensor is responsible for the radial transport of angular momentum through the disk, while the $z\phi$ component transports angular momentum vertically via the disk wind. Specifically, the local mass accretion rate through the disk is set by sum of the vertically integrated radial stress and the wind stress evaluated on a constant $\theta$ disk surface (see equation 22 of \citealt{2013ApJ...769...76B} for how the Maxwell stress tensor relates to disk accretion). The $R\phi$ stress is largest in the midplane current layer, however, $T_{z\phi}$ quickly becomes larger than $T_{R\phi}$ moving away from the midplane, and reaches its maximum magnitude at less than half a scale height away from the midplane. At the initially defined disk surface of two scale heights near the base of the wind, $T_{z\phi}$ is still an order of magnitude larger than $T_{R\phi}$. Thus, angular momentum is transported vertically away from the disk midplane by the magnetic wind stress, leading to the rapid accretion in the current layer and both the outward expanding upper regions of the disk and the launching of the disk wind (see the green dotted line in Fig.~\ref{fig:stress}).

\subsection{The formation of rings and gaps in the disk}\label{sec:ringgap}

Central to the formation of rings and gaps is the thin midplane current sheet. The creation of the current sheet starts near the upper and lower surfaces of the disk as the differential rotation between the disk and the initially static disk corona rapidly induces a toroidal magnetic field there. The toroidal magnetic field is responsible for driving an outward (positive) radial current, $J_r \approx \frac{c}{4\pi} \frac{dB_\phi}{rd\theta}$, in the disk since $B_\phi$ increases from negative to positive as the polar angle $\theta$ increases from above to below the disk midplane. The resulting Lorentz force, $F_{L,\theta}\propto J_r B_\phi$ (associated with the magnetic pressure gradient in the $\theta$-direction), pushes the ions (and the toroidal magnetic field lines tied to the ions) towards the midplane, thus moving the surface current layers closer together until they combine at the disk midplane to form a single current sheet where $B_\phi$ reverses polarity. The magnetic pressure gradient is further steepened at the magnetic null due to the effects of AD \citep{1994ApJ...427L..91B}, creating a thin midplane current layer. Finally, the Lorentz force exerted due to the radial current ($F_{L,\phi}\propto J_r B_z$) is in the $-\phi$ direction, draining angular momentum in the current layer leading to strong accretion and the inward pinch of the poloidal magnetic field there. The eventual reconnection of the radial magnetic field as it is dragged inward leads to the creation of a poloidal magnetic field loop, thereby leaving the region that it encloses devoid of vertical (or poloidal) magnetic flux. In neighboring regions, however, the poloidal magnetic field concentrates. The disk mass then grows in the regions where the vertical magnetic flux is lowest (rings) and the neighboring regions with larger poloidal magnetic flux can drive faster accretion through them. This phenomenon is the essence of the reconnection-driven ring and gap formation mechanism described in Section 3.3 of \citetalias{2018MNRAS.477.1239S} based on 2D axisymmetric simulations (see their figures 6 and 7). It is also apparent in our 3D reference simulation, as illustrated in Fig.~\ref{fig:panels}(c), which shows the eventual magnetic field configuration of alternating bands of toroidal and poloidal magnetic field concentrations and the radial pinch in the magnetic flux contours at the current layer where $B_\phi=0$.

\begin{figure}
\centering
\includegraphics[width=1.0\columnwidth]{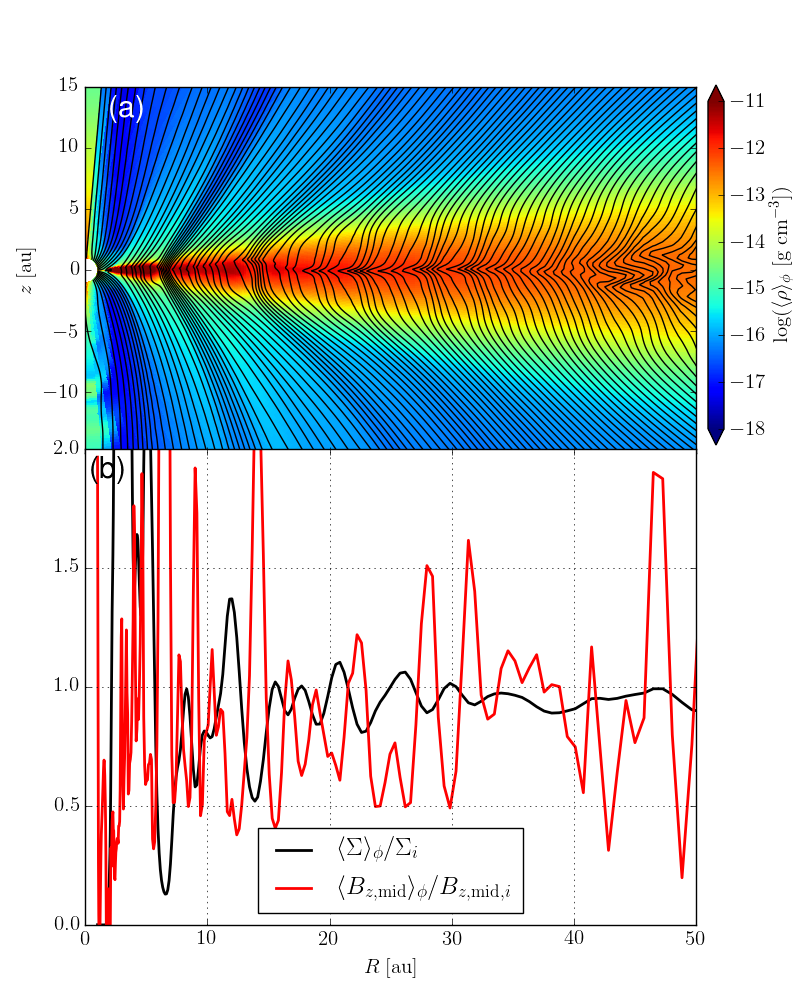}
\caption{The reference simulation at a representative time $t/t_0=1500$. The top panel plots the logarithm of the density (colour map) and the $\phi$ integrated magnetic flux contours (or effective poloidal field lines, black lines). The bottom panel shows the $\phi$ averaged surface density (black) and vertical magnetic field at the disk midplane (red) normalized respectively by their initial distribution. (See the supplementary material in the online journal for an animated version of panel (a) of this figure.)}
\label{fig:bzmid}
\end{figure}

The end result of the reconnection of a radially pinched magnetic field is the redistribution of the poloidal magnetic flux relative to the disk material, as illustrated in Fig.~\ref{fig:bzmid}. The top panel shows the $\phi$-averaged density distribution (colour map) and the effective poloidal field lines (black lines). It is clear that the poloidal field lines are distributed very unevenly, especially inside the disk. Specifically, the field lines are concentrated in some regions but spread out in others. The corresponding strong variation of the vertical magnetic field strength at the disk midplane is quantified in Fig.~\ref{fig:bzmid}(b) (red line). For comparison, the surface density distribution is also plotted in the same panel (black line). It is clear that the surface density is strongly anti-correlated with the vertical field strength, especially in the region around 10~au, where the contrast between the dense, weakly magnetized rings and the more diffuse but more strongly magnetized gaps is the largest. This anti-correlation starts at early times when the rings and gaps are still nearly axisymmetric, so it is likely created by the same mechanism as in the 2D (axisymmetric) simulations, namely, the redistribution of poloidal magnetic flux relative to disk material through reconnection. The anti-correlation persists to later times (such as that shown in Fig.~\ref{fig:bzmid}) when the rings and gaps become more non-axisymmetric (see the next subsection). As the resolution of this 3D simulation is $\sim 30\%$ lower than that of the 2D simulations in \citetalias{2018MNRAS.477.1239S} and the effective poloidal magnetic field lines are drawn by integrating over the azimuthal angle, it is more difficult to observe the field lines in the act of reconnecting (as in figure~7 of \citetalias{2018MNRAS.477.1239S}). Nonetheless, the close similarity between the 2D and 3D simulations, especially in the severe radial pinching of the poloidal magnetic field in a thin fast-accreting layer and the anti-correlation between the disk surface density and poloidal field strength, leaves little doubt that the same mechanism is at work in both 2D and 3D. We should note that this is not the only mechanism involving magnetic fields that can create rings and gaps; such substructures are observed in shearing box simulations without a prominent midplane current sheet (e.g., \citealt{2012A&A...548A..76M}).

\subsection{Non-axisymmetric disk substructure}\label{sec:phivar}

\subsubsection{Magnetic fields}

\begin{figure*}
\centering
\includegraphics[width=2.0\columnwidth]{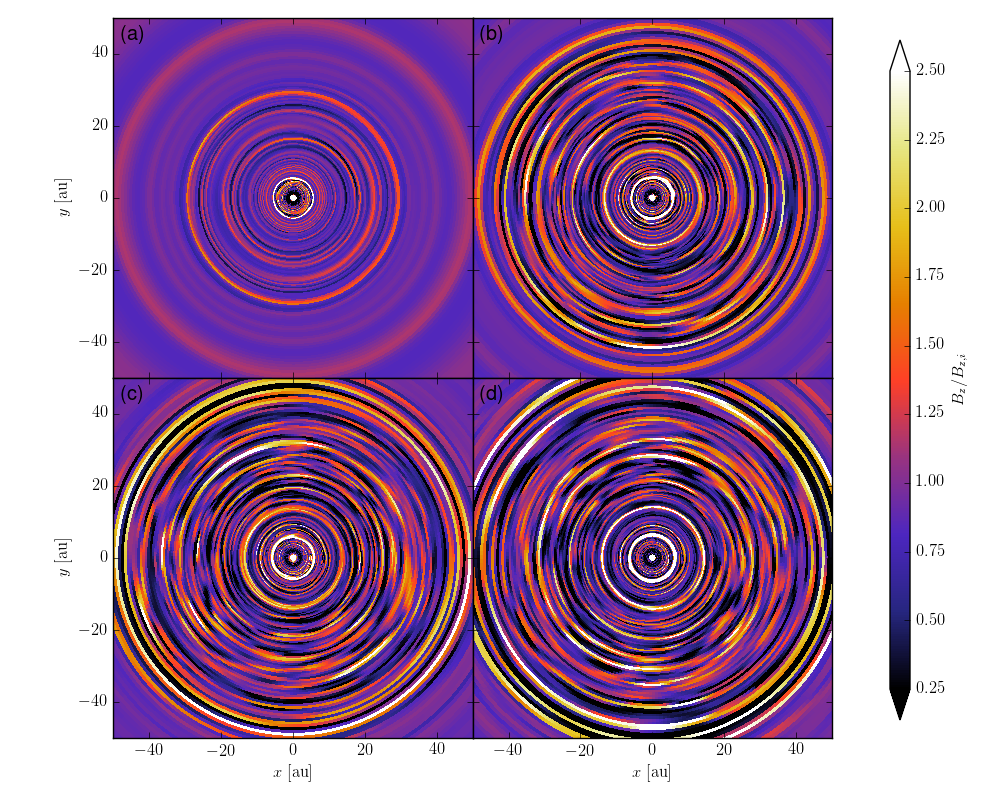}
\caption{Spatial variation of the vertical magnetic field strength at the midplane normalized to its initial value for the reference simulation at four representative times: (a) $t/t_0=500$, (b) $1000$, (c) $1250$, and (d) $1500$. They show clearly that the initially axisymmetric rings of enhanced vertical field are perturbed but not disrupted in 3D. (See the supplementary material in the online journal for an animated version of this figure.)}
\label{fig:bz4panels}
\end{figure*}

The vertical magnetic field strength at the midplane, $B_{z,\mathrm{mid}}$, has prominent {\it radial} substructure as illustrated in Fig.~\ref{fig:bzmid}(b). Substantial substructure in $B_{z,\mathrm{mid}}$ develops in the {\it azimuthal} direction as well, as shown in Fig.~\ref{fig:bz4panels}. Since the disk is initially assumed to be axisymmetric, it is not surprising that the rings and gaps remain relatively axisymmetric at early times when their amplitudes are still low, as in panel (a) of Fig.~\ref{fig:bz4panels} ($t/t_0=500$). By the time shown in panel (b) ($t/t_0=1000$), the rings and gaps have grown much more prominent. Many of these (magnetic) rings have also become significantly non-axisymmetric, with some parts brighter than others. The azimuthal variation is particularly striking in the intermediate range of radii, between $\sim15$ and 40~au, where some of the rings appear to be frayed into multiple threads. The frayed ringlets occasionally merge together, creating wider structures. The fraying and merging continue at later times, creating rich non-axisymmetric substructures that are illustrated in panel (c) at $t/t_0=1250$ and in panel (d) at $t/t_0=1500$. The formation and evolution of the substructures can be seen more clearly in the supplementary movie online, which shows the occasional development of short spurs on the rings that are subsequently sheared into flocculent spirals by differential rotation.

How the non-axisymmetric substructures are created is unclear. Presumably the azimuthal variation develops from the numerical noise associated with the finite simulation grid since the initial conditions are axisymmetric and there is no explicit perturbation added. It may be amplified by several potential instabilities in the system, both inside the disk and in the wind. First, the amplification could be due to the development of non-axisymmetric modes of the MRI \citep{1992ApJ...400..610B}, particularly at relatively large radii where the ambipolar Elsasser number is larger and axisymmetric (`channel flow') modes are clearly visible (see, e.g., Fig.~\ref{fig:global}). Second, it could be caused by the wind, which is dominated by a toroidal magnetic field (see Fig.~\ref{fig:panels}c and Fig.~\ref{fig:alongBfield}d) that is prone to kink instabilities (e.g., \citealt{2006ApJ...653L..33A}). Third, there is a substantial radial velocity difference between the midplane accretion layer and the rest of the disk, which could be prone to Kelvin-Helmholtz instability, as it is thought to play a role in turning the axisymmetric `channel flows' in 2D into a more turbulent (non-axisymmetric) flow in 3D \citep{1994ApJ...432..213G,1995ApJ...440..742H}. Finally, since there is a sharp reversal of the toroidal magnetic field across the accretion layer, it is possible that reconnection of the sharply pinched toroidal field plays a role in producing the prominent azimuthal variations shown in Fig.~\ref{fig:bz4panels}. The last possibility is particularly intriguing, because the sharp reversal of toroidal magnetic field lies at the heart of the proposed mechanism for the formation of rings and gaps in 2D.

\begin{figure}
\centering
\includegraphics[width=1.0\columnwidth]{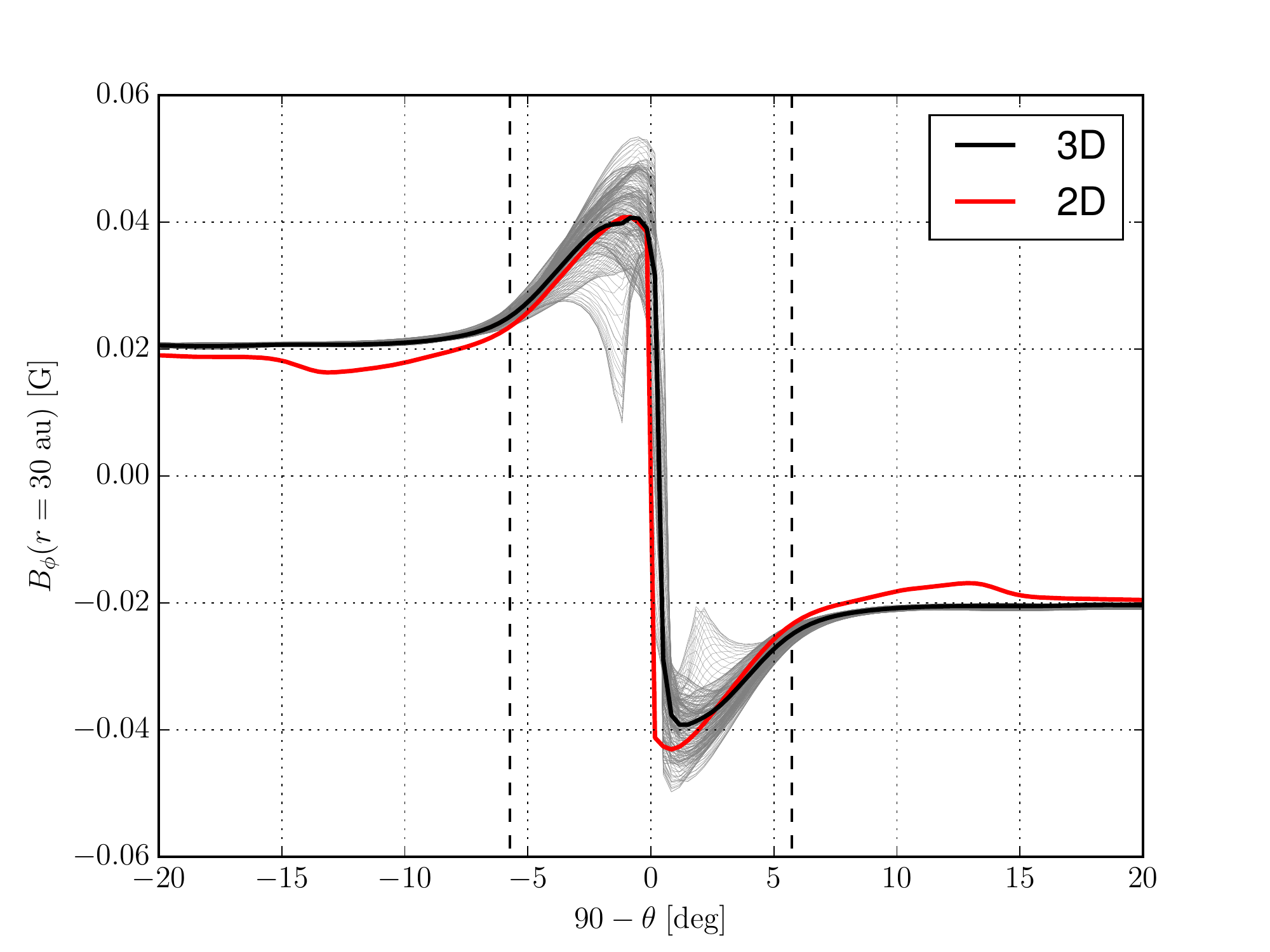}
\caption{The variation of the toroidal component of the magnetic field $B_\phi$ as a function of $90-\theta$~[deg] (negative is below the midplane and positive is above the midplane) at a representative radius $r=30$~au in the reference 3D simulation ($\phi$-averaged) and the corresponding 2D simulation at $t/t_0=1500$. The thin grey lines plot $B_\phi$ at all $\phi$.}
\label{fig:bphi}
\end{figure}

\begin{figure}
\centering
\includegraphics[width=1.0\columnwidth]{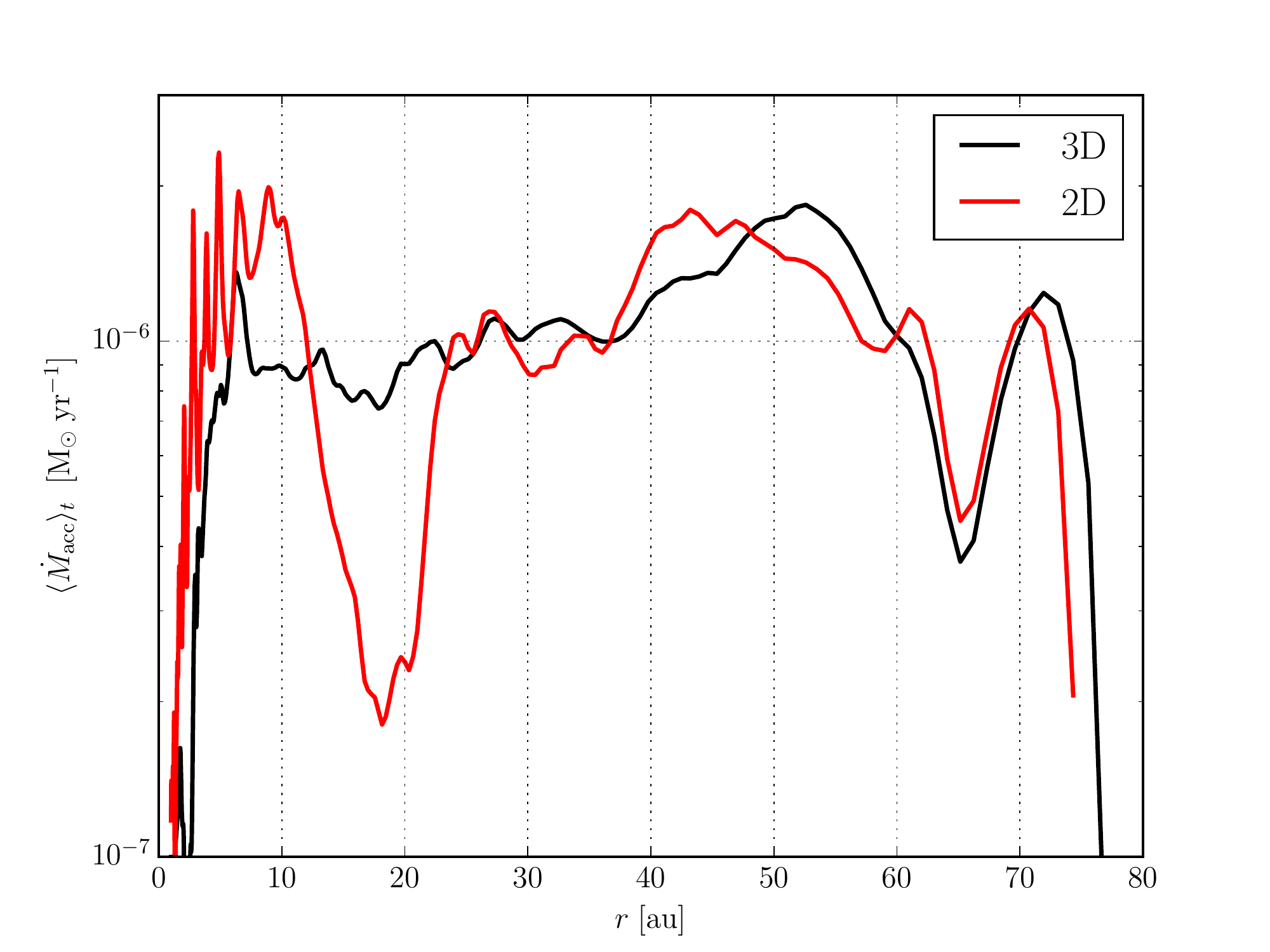}
\caption{The time averaged mass accretion rate from $t/t_0=1000$ to 1500 as a function of disk radius for both the 2D and 3D simulations. The broad similarity between the two means that the pinching of the magnetic field lines in the azimuthal direction that drives the mass accretion is not significantly reduced, if at all, by reconnection in 3D.}
\label{fig:mdot_tavg}
\end{figure}

One concern is that the reconnection of the toroidal magnetic field in 3D, if efficient enough, may shut down the ring and gap formation mechanism identified in 2D altogether. This is because reconnection of the toroidal field can erase the radial component of the current density $J_r\propto \partial B_\phi/\partial \theta$ and thus weaken the magnetic torque ($\propto J_r B_\theta$) that drives the fast midplane accretion. However, as mentioned earlier, Fig.~\ref{fig:panels}(c) and Fig.~\ref{fig:alongBfield}(d) demonstrate explicitly that the toroidal field $B_\phi$ in the 3D simulation still changes sign sharply over a short vertical distance and Fig.~\ref{fig:panels}(d) shows that fast midplane accretion still exists and can drag the poloidal field lines into a sharply pinched configuration prone to reconnection (see also Fig.~\ref{fig:global}). Indeed, the vertical gradient of the toroidal field near the midplane in the 3D simulation is just as sharp as, if not sharper than, that in its 2D counterpart, as shown in Fig.~\ref{fig:bphi}. The mass accretion rate in the 3D simulation is also comparable to that in its 2D counterpart over most of the disk (Fig.~\ref{fig:mdot_tavg}), indicating that reconnection of $B_\phi$, if present, does not weaken the magnetic torque that drives the significant disk accretion. These broad similarities indicate that, although reconnection of the toroidal magnetic field is expected in 3D and may play a role in the development of non-axisymmetric substructure, it does not suppress the mechanism of ring and gap formation through the radial reconnection of the poloidal field lines completely. The end result is that the prominent rings and gaps in the distribution of the midplane vertical magnetic strength $B_{z,\mathrm{mid}}$ in axisymmetric simulations are strongly perturbed but not completely disrupted in 3D (see Fig.~\ref{fig:bz4panels}c,d and especially Fig.~\ref{fig:bzmid}b).

\begin{figure*}
\centering
\includegraphics[width=2.0\columnwidth]{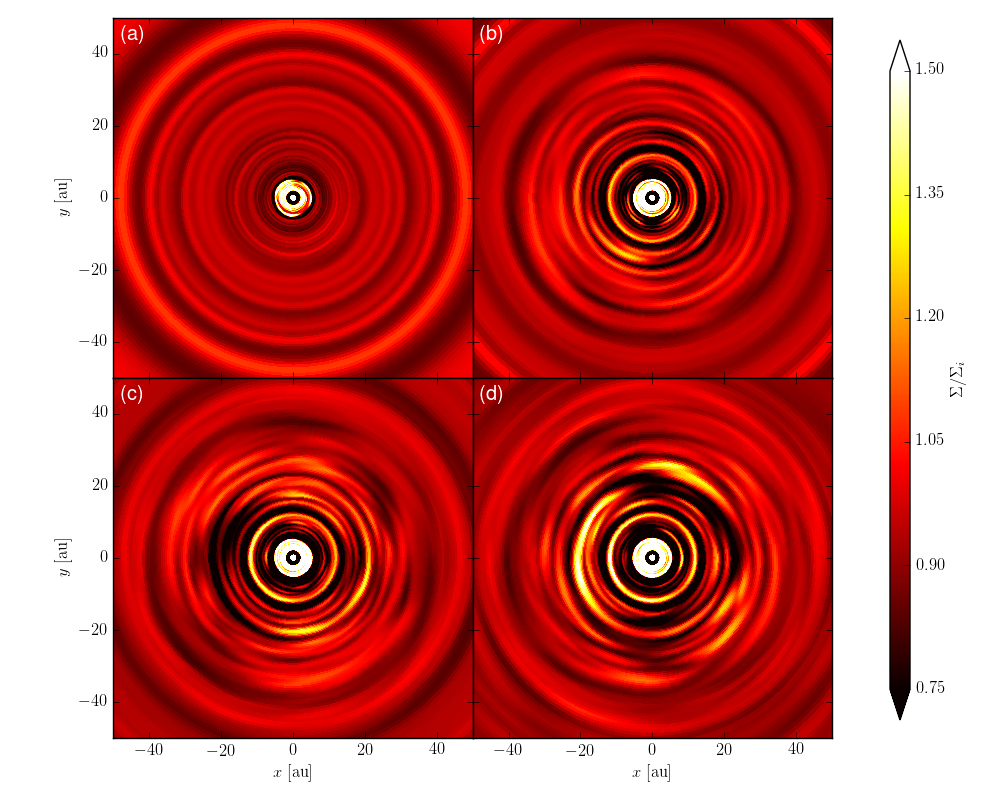}
\caption{Spatial variation of the disk surface density normalized to its initial value for the reference simulation at four representative times: (a) $t/t_0=500$, (b) $1000$, (c) $1250$, and (d) $1500$. They show clearly that the initially axisymmetric rings of enhanced surface density are perturbed but not disrupted in 3D. (See the supplementary material in the online journal for an animated version of this figure.)}
\label{fig:surfden4panels}
\end{figure*}

\subsubsection{Surface density}

Azimuthal variation develops in the surface density distribution as well, as illustrated in Fig.~\ref{fig:surfden4panels}, which plots the surface density distributions for the same four representative times as in Fig.~\ref{fig:bz4panels}. Several features stand out in the figure. First, there is a plateau of enhanced surface density (relative to the initial distribution) near the center within a radius of $\sim 5$~au that is apparent in all four panels. It corresponds to the region where repeated reconnection of the poloidal magnetic field is clearly visible even after azimuthal averaging. The reconnection weakens the poloidal magnetic field relative to the matter (see Fig.~\ref{fig:bzmid}b), making it easier for the mass to accumulate there. This behavior is consistent with the proposed mechanism of dense ring formation through reconnection of the poloidal field lines, although this region may be affected by its proximity to the inner boundary.

The (bright) plateau is surrounded by a prominent (dark) gap at a radius of approximately 6~au, where the surface density is depleted relatively to the initial value by almost an order of magnitude at the end of the simulation (see Fig.~\ref{fig:bzmid}b, black line). Inside this low surface density gap, there is a local bunching of the poloidal magnetic field lines, as shown in Fig.~\ref{fig:bzmid}a, which leads to an increase of the poloidal field strength well above its initial value (Fig.~\ref{fig:bzmid}b, red line). It corresponds to the bright nearly axisymmetric ring at $r\approx 6$~au in the distribution of the midplane vertical field strength seen in Fig.~\ref{fig:bz4panels}. The same pattern repeats itself at later times for a larger radius of $\approx 15$~au, where there is again a persistent axisymmetric gap of depleted surface density (Fig.~\ref{fig:surfden4panels}b,c,d) coincident with an axisymmetric ring of enhanced poloidal magnetic field (Fig.~\ref{fig:bz4panels}b,c,d). Just interior to this low-surface density gap, there is a bright ring of enhanced surface density where the poloidal field strength is much weaker than in its neighboring gap. The persistence of these complete, nearly axisymmetric, rings and gaps (in the distributions of both surface density and the vertical field strength at the disk midplane) over tens to hundreds of local orbital periods is strong evidence for the stability of such substructures in 3D, at least in the inner disk region of the reference simulation (within $\sim 15$~au).

In the disk region of intermediate radius range between $\sim 15$ and $40$~au, the rings and gaps start out nearly axisymmetric (see panel a of Fig.~\ref{fig:surfden4panels}), but become more clumpy and distorted at later times (see panels b-d). In particular, a well-defined bright clump (of enhanced surface density) starts to develop out of a fainter arc on the ring at a radius of $\sim 18$~au around $t/t_0=700$. It completes approximately three orbits before fading away into a faint arc that is still visible at the time $t/t_0=1000$ shown in panel (b), (the arc at approximately 7 o'clock angle). The formation and disappearance of the bright clump can be seen most clearly in the movie of the disk surface density distribution in the online journal. Although exactly why this clump forms is unclear, it could be related to the Rossby wave instability (e.g., \citealt{1999ApJ...513..805L,2015arXiv150906382A}). In any case, the clump saturates at a relatively moderate amplitude before decaying away, and its formation perturbs but does not disrupt the parent ring. As is the case for the `magnetic rings' shown in Fig.~\ref{fig:bz4panels}, some of the gas rings fray into multiple thinner ringlets, while others merge into wider structures. The dynamic evolution of the (significantly non-axisymmetric) surface density structures in this region mirrors that in the midplane vertical magnetic field $B_{z,\mathrm{mid}}$. In particular, the flocculent spirals that were already visible in the $B_{z,\mathrm{mid}}$ distribution become more prominent. Indeed, there is a strong anti-correlation between the surface density and $B_{z,\mathrm{mid}}$ all over the disk, as previously discussed in Section~\ref{sec:ringgap}.

\begin{figure*}
\centering
\includegraphics[width=1.0\columnwidth]{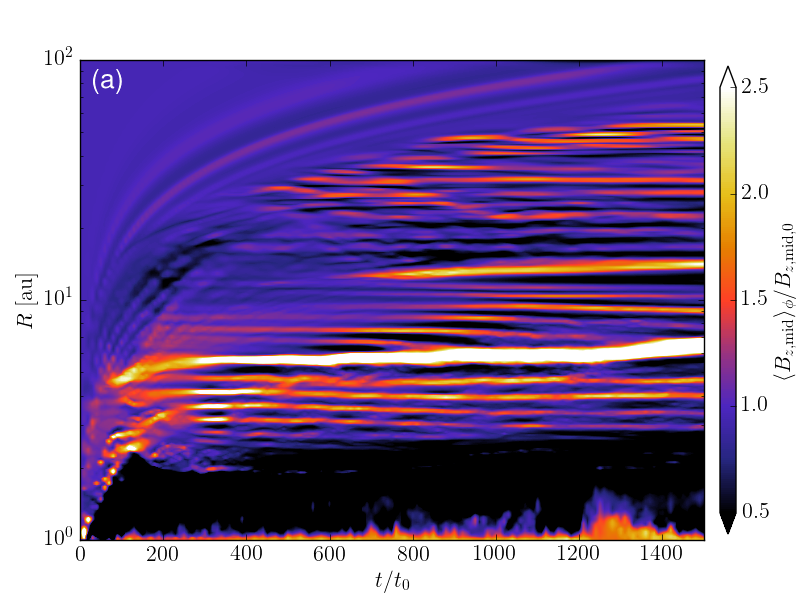}
\includegraphics[width=1.0\columnwidth]{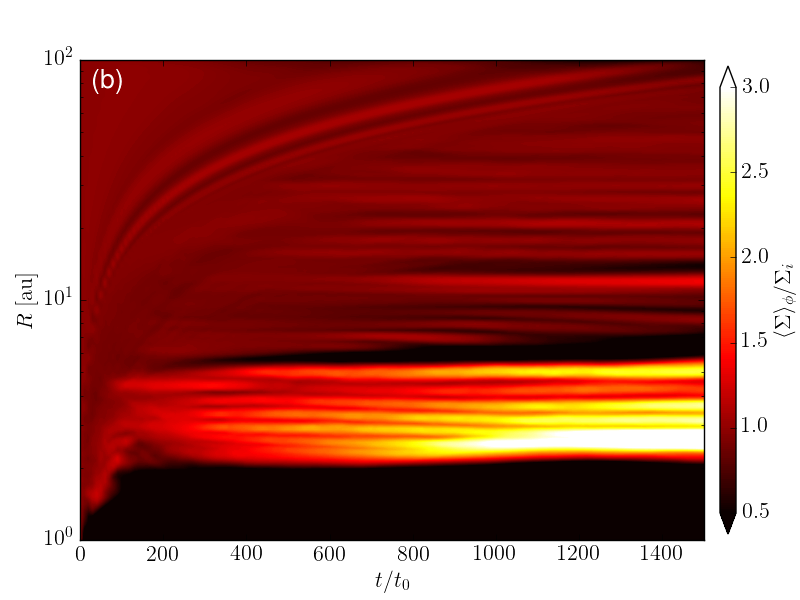}
\includegraphics[width=1.0\columnwidth]{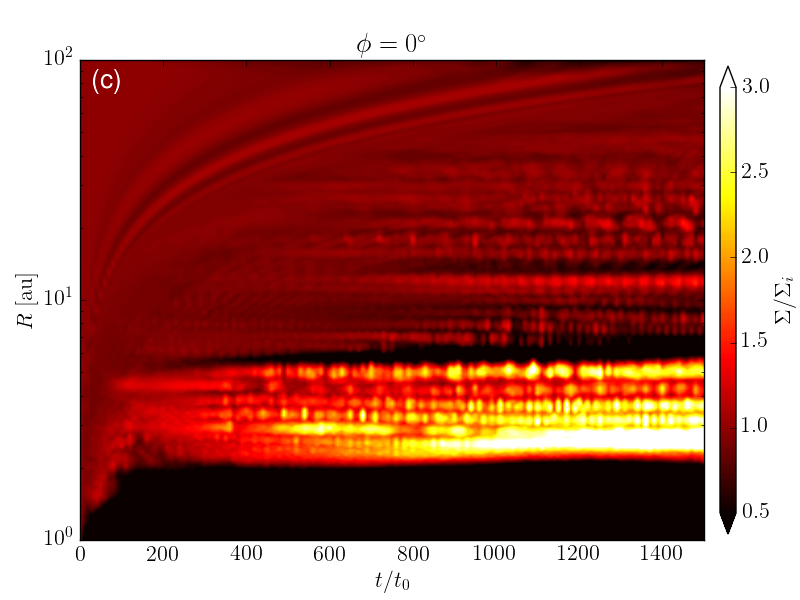} \includegraphics[width=1.0\columnwidth]{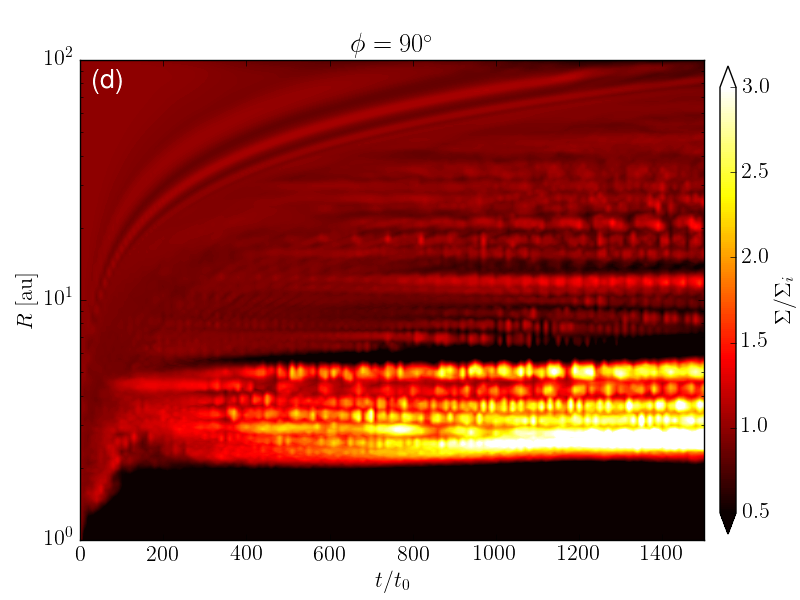}
\caption{The color contours show the azimuthally averaged (a) vertical magnetic field strength at the disk midplane and (b) surface density, and the surface density at two representative azimuthal directions, (c) $\phi=0^\circ$ and (d) $\phi=90^\circ$, all as a function of radius and time. Both the vertical magnetic field and surface density are normalized to their initial radial distributions.}
\label{fig:space-time}
\end{figure*}

To further illustrate the longevity and coherency of the disk substructures, Fig.~\ref{fig:space-time} plots the space-time diagrams for the azimuthally averaged vertical magnetic field at the disk midplane, the azimuthally averaged surface density, and the surface density along two representative azimuthal directions ($\phi=0^\circ$ and $90^\circ$), all normalized to their initial radial distributions. Panel (a) shows that, although there is some evidence for merging (e.g., near $\sim 7$~au), the `magnetic rings' and gaps stay relatively well-defined in both space and time. This is also true for the azimuthally averaged gas rings and gaps, most of which remain distinct throughout the simulation (panel b), despite the fact that there is substantial time variation in such substructures along a given azimuthal direction (panels c and d). The disk substructures are long-lived enough (at least $\sim1000$ inner orbital periods) to be effective as potential dust traps.
 
\section{Parameter survey}\label{sec:param}

We have carried out five simulations in addition to the reference run (Table~\ref{tab:sims3D}). Model ad-els0.25 is a repeat of the reference simulation but with a lower resolution ($200\times180\times180$ instead of $300\times270\times270$). It is done to both explore the resolution effects and for direct comparison with simulations with other physical parameters, which are carried out at a lower resolution compared with the reference simulation because of constraints on computational resources. It will sometimes be referred to as `the low-res reference run' below. Model ad-els0.05 is five times more diffusive than the low-res reference run (with a characteristic AD Elsasser number $\Lambda_0=0.05$ instead of 0.25), Model ad-els1.25 is five times less diffusive (or better magnetically coupled), and Model ad-els$\infty$ does not have any ambipolar diffusion at all (i.e., ideal MHD). In addition, Model beta1e4 has a weaker initial magnetic field than the low-res reference run, with an initial midplane plasma-$\beta$ of $10^4$ rather than $10^3$. We will start with a discussion of the low-res reference simulation, which serves a benchmark against which other simulations of the same resolution are compared.

\begin{figure*}
\centering
\includegraphics[width=2.0\columnwidth]{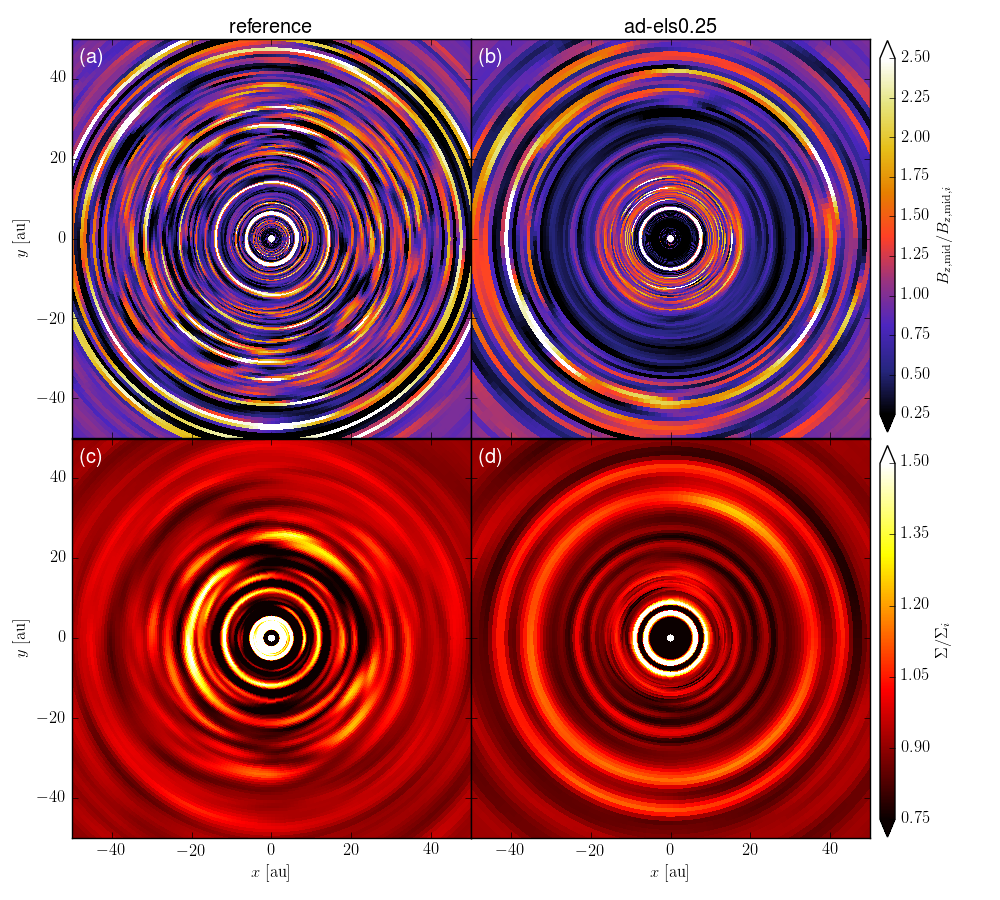}
\caption{Spatial variation of the vertical magnetic field strength at the midplane (top row) and the disk surface density (bottom row) normalized to their respective initial distribution for the high-resolution reference simulation (left column) and its low-resolution counterpart (model ad-els0.25, right column) at a representative time $t/t_0=1500$.}
\label{fig:surfden4panels-hires}
\end{figure*}

\subsection{Low-resolution reference run}\label{subsec:res}

The low-res reference run is broadly similar to the reference run. It forms both magnetic ($B_{z,\mathrm{mid}}$) and gas rings and gaps that are nearly axisymmetric at early times. Non-axisymmetric variations, including fraying/merging of the rings and clumping within rings, develop at later times but saturate at a relatively moderate amplitude such that the rings and gaps become strongly perturbed but not disrupted. There is a strong anti-correlation between the poloidal magnetic field and the disk surface density, as in the reference case. The main difference is that it takes longer for the non-axisymmetric variations to develop in the lower resolution model compared to the reference model. This difference is illustrated in Fig.~\ref{fig:surfden4panels-hires} for a representative time $t/t_0=1500$. It is clear that the higher resolution model has far more structures in the distributions of both the midplane vertical magnetic field and the surface density, especially in the middle radius range between $\sim 15$ and 40~au. This is the region where, in the reference model, the poloidal field lines are most sharply pinched near the midplane and the toroidal field reverses its direction quickly over a short vertical distance, as illustrated in Fig.~\ref{fig:bzmid}a and Fig.~\ref{fig:bphi}, respectively. The lower resolution in Model ad-els0.25 may have reduced the sharpness of the field pinching, and therefore the likelihood of reconnection, in both the radial and azimuthal direction, which are the main drivers of the radial structures (rings and gaps) and possibly the azimuthal variations, respectively. Another difference is that the lower resolution simulation has a larger region of a `dipole-like' closed poloidal magnetic field that extends several au beyond the inner boundary; the disk dynamics in this region may be affected by how the boundary is treated. The advantage of the lower resolution is that it allows the simulation to reach a later times (3000~$t_0$ versus 1500~$t_0$). As we will show in the next subsection, more substructures do develop at later times in the lower resolution reference model, making it more closely resemble the higher resolution reference model. We conclude that the development of prominent disk substructures is a robust feature of the coupled disk-wind system, although their detailed properties are resolution-dependent.

\subsection{Strength of magnetic coupling}\label{subsec:ad}

\begin{figure*}
\centering
\includegraphics[width=2.0\columnwidth]{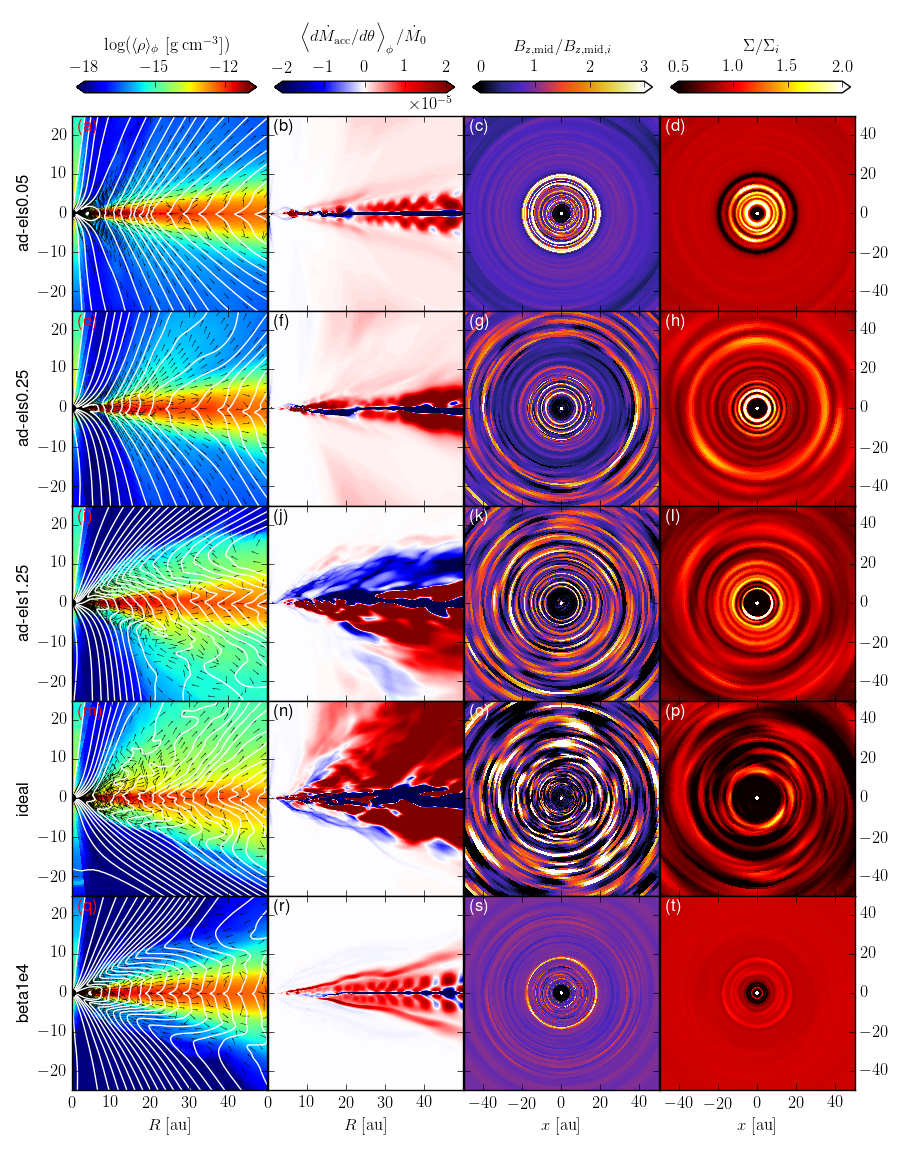}
\caption{The simulations from the top row to the bottom row: ad-els0.05, ad-els0.25, ad-els1.25, ideal, beta1e4 at $t/t_0=2500$. Plotted are azimuthally averaged density and poloidal field lines (first column), mass accretion rate per unit polar angle (second column), normalized midplane vertical field strength (third column) and surface density (fourth column).}
\label{fig:5x4}
\end{figure*}

The effects of different levels of ambipolar diffusion are illustrated in Fig.~\ref{fig:5x4}. The left two columns of Fig.~\ref{fig:5x4} show the `edge-on' plots of the azimuthally-averaged poloidal density and magnetic field lines and the mass accretion rate per unit polar angle. The right two columns show the `face-on' views of the vertical magnetic field strength at the disk midplane and gas surface density. The panels in Fig.~\ref{fig:5x4} are plotted at a common representative time of $t/t_0=2500$ for all five of the low-resolution models. The first impression is that systems with different levels of ambipolar diffusion have the following in common: they all are actively accreting with outflows of various types, and substantial substructures develop in all disks. There are, however, significant differences. In the most diffusive case (model ad-els0.05), both the disk and the wind are largely laminar and mirror-symmetric above and below the midplane (panel a), with the disk accretion confined to the midplane (panel b). Prominent rings and gaps form in both the distributions of the midplane vertical field (panel c) and surface density (d) within a radius of about 20~au. They remain nearly axisymmetric until the end of the simulation ($t/t_0=3000$). As the ambipolar diffusivity decreases by a factor of five to the reference value used in model ad-els0.25, the wind and disk remain largely laminar, although the accretion layer has migrated close to the bottom surface of the disk within a radius of about 20~au (panel f), leads to a significant asymmetry about the disk midplane at $\theta=90^\circ$. Nevertheless, complete (as opposed to partial) rings and gaps that close on themselves are still formed in this region in the distributions of both midplane vertical field (panel g) and surface density (panel h). At larger radii, the rings and gaps appear somewhat less well-defined, especially in the distribution of midplane vertical field, with some rings containing multiple strands and others appearing as tightly wound spirals (panel g). This contrast in appearance between rings and gaps in the inner and outer parts of the disk becomes more pronounced as the ambipolar diffusivity decreases by another factor of five to the value used in model ad-els1.25. In this better magnetically coupled system, the motions of both the disk and outflow become more chaotic (panels i and j). In particular, the mass accretion is no longer confined to the midplane region (panel j). As in the well-coupled 2D (axisymmetric) simulations \citepalias{2018MNRAS.477.1239S}, highly variable `avalanche' accretion streams develop near the disk surface, which dominate the dynamics in the disk envelope (e.g., \citealt{2018ApJ...857...34Z}). In particular, they induce a strong asymmetry in the top and bottom half of the disk envelope, with rapid accretion in the top half (reminiscent of the funnel-wall accretion; \citealt{2018ApJ...857....4T}) but rapid expansion in bottom half (panel i). Nevertheless, rather regular substructures are still formed in the disk surface density, especially in the inner part (within about 20~au), where well-behaved rings and gaps are still evident (panels k and l). At larger radii, the substructures appear more spiral-like, especially in the surface density distribution (panel l); the spirals can be seen even more clearly in the animation of the figure in the online journal. The difference in appearance between the inner and outer parts of the disk is broadly similar to that in the more magnetically diffusive model ad-els0.25, but the outer disk of the less diffusive model exhibits a more spiral-like appearance.

Even more pronounced spirals develop in the completely coupled (ideal MHD) model ad-els$\infty$. The gas motions are chaotic in both the disk and the envelope (panels m and n), driven presumably by the MRI in the disk and its variants (`avalanche accretion streams') in the envelope. As in the moderately well-coupled case of model ad-els1.25, there is a strong asymmetry in the top and bottom half of the simulation, which is unsurprising given the rather turbulent state of the system. The ideal MHD disk is full of substructures (panels m and n), but they are less coherent than those in model ad-els1.25. In particular, the substructures in the distribution of the midplane vertical magnetic field appear more clumpy, with most of the `blobs' of enhanced $B_{z,\mathrm{mid}}$ smeared into arcs rather than complete rings (panel o). The difference in the disk surface density substructure is even more striking: they are dominated by flocculent spirals rather than rings and gaps (panel p). The spirals can be seen in the distribution of not only surface density, but also the vertical magnetic field at the disk midplane. They are common in global ideal MHD simulations (e.g., figure~9 of \citealt{2000ApJ...528..462H}, figure~10 of \citealt{2011ApJ...735..122F}, figure~7 of \citealt{2014ApJ...784..121S}).

A clear trend emerges from this set of simulations: as the ambipolar diffusivity increases, both the disk and wind become more laminar, and the disk substructures become more coherent spatially (i.e., more ring- and gap-like). This trend is perhaps to be expected. In the completely coupled (ideal MHD) limit, the disk is expected to be unstable to both axisymmetric (channel) and non-axisymmetric MRI modes. The latter should make it easier to form substructures of limited azimuthal extent (or clumps) than complete rings. It is also possible that the poloidal magnetic field lines are wrapped more strongly by the vertical differential rotation in the ideal MHD limit, creating a more severe pinching of the field lines in the azimuthal direction. This would make the toroidal magnetic field more prone to reconnection, which is necessarily non-axisymmetric and likely leads to clump formation. The shearing of the clumps by differential rotation, coupled with relatively fast accretion\footnote{The fast accretion in the ideal MHD case can be inferred from Fig.~\ref{fig:5x4}(p), which shows a lower surface density (and thus more mass depletion) than the more magnetically diffusive cases.}, would lead to widespread spirals, as seen in model ad-els$\infty$. As the ambipolar diffusivity increases, the MRI (and its variant `avalanche accretion streams') should be weakened and eventually suppressed. In the limit where the MRI is suppressed, active accretion is still possible provided that the ambipolar diffusivity is not too large (there would not be any accretion and structure formation if the magnetic field is completely decoupled from the matter). Accretion is confined to the disk midplane region, driven by angular momentum removal from the largely laminar disk wind (e.g., Fig.~\ref{fig:5x4}a,b) and aided by the steepening of the radial ($J_r$) current sheet through ambipolar diffusion via the Brandenburg-Zweibel mechanism  \citep{1994ApJ...427L..91B}. The midplane accretion would drag the poloidal field lines into a sharply pinched configuration, leading to reconnection and the eventual formation of rings and gaps through the mechanism discussed earlier (and in \citetalias{2018MNRAS.477.1239S}). Once formed, such substructures can be long lived because they are not prone to disruption by the non-axisymmetric modes of the MRI, which are suppressed by ambipolar diffusion in this limit. Azimuthal variation is still possible because of, e.g., reconnection of the oppositely directing toroidal magnetic field across the accretion layer. Again, this reconnection is intrinsically non-axisymmetric but appears to only perturb rather than destroy the rings and gaps. 

At a more fundamental level, it is perhaps not too surprising that the magnetic diffusivity can make the disk substructures more ordered and coherent. Magnetic diffusivity is expected to reduce disorder in the magnetic field and a more ordered magnetic field configuration is ultimately responsible for the creation of more coherent disk substructures such as rings and gaps. This notion is consistent with the radially demarcated pattern observed in the moderately well coupled model ad-els1.25, where well-defined rings and gaps are produced in the more-diffusive inner disk (as measured by the dimensionless Elsasser number, which scales with radius as $\Lambda \propto r^{3/4}$ initially) and flocculent spirals in the better-coupled outer disk. It is also consistent with the trend that, as the ambipolar diffusivity increases from model ad-els1.25 to model ad-els0.25, rings and gaps become more important relative to flocculent spirals (compare panels h and l of Fig.~\ref{fig:5x4}). 

Similarly, one might also expect that the ring and gap spacing produced from the reconnection of the poloidal magnetic field at the midplane current sheet will be set by the diffusive processes in the disk. For example, in the Sweet-Parker theory of slow reconnection (\citealt{1957JGR....62..509P,1958IAUS....6..123S}; see also the review by \citealt{2009ARA&A..47..291Z}), the length scale over which the magnetic field will reconnect is $L_\mathrm{rec}\propto S^{1/2}$, where the Lundquist number is $S\sim hv_A/\eta$. Thus, in the less diffusive outer disk regions of these simulations,\footnote{For example, the reference simulation is initialized such that the Elsasser number is greater than unity for radii $r\gtrsim 6.3$~au at the disk midplane ($\theta=\pi/2$) and for radii $r\gtrsim 21$~au at the disk surface ($\theta=\pi/2\pm\theta_0)$, before $\rho_i$ increases rapidly towards the poles and $\Lambda\rightarrow\infty$.} we should expect to see a general trend of the ring and gap separations increasing as a function of radius. This seems to be the case as seen in the right two columns of Fig.~\ref{fig:5x4}, but the result may be complicated by the logarithmic radial grid spacing, and it is difficult to draw any firm conclusions from the simulations with varying Elsasser numbers. Further investigation is required to evaluate more thoroughly the conjecture that the ring and gap spacing is controlled by the detailed microphysics of the non-ideal MHD effects.

\subsection{Magnetic field strength}\label{subsec:beta}

Finally, we explore briefly the effects of the initial magnetic field strength with model beta1e4, which has the same physical parameters and resolution as the low-res reference model ad-els0.25 except for an initial plasma-$\beta$ of $10^4$ instead of $10^3$. Unlike its more strongly magnetized counterpart, the magnetic structure above and below the disk is strongly affected by accretion streams (panel q of Fig.~\ref{fig:5x4}). Nevertheless, the dynamics inside the disk appear more similar, with the accretion concentrating near the midplane and expansion outside the thin accretion layer (panel r). Rings and gaps are still formed in both vertical field strength and surface density, but they are far less prominent than their counterparts in the more strongly magnetized reference model (compare panels s and g and panels t and h). The weaker substructure in $B_{z,\mathrm{mid}}$ is likely because the magnetic field strength is too weak to drive midplane accretion vigorous enough to pinch the poloidal field lines to the point of reconnection. The latter comes about because it is harder for a weaker field to move the gas around and, therefore, to create prominent substructure in the disk surface density. Nevertheless, there is still an anti-correlation between the surface density and vertical field strength at the disk midplane, indicating that the same mechanism for ring and gap formation is at work even in the more weakly magnetized case.

\section{Discussion}\label{sec:discuss}

\subsection{Poloidal magnetic fields and magnetic diffusivity-regulated disk accretion}

One of the key assumptions of all but one of the simulations presented in the paper is that the disk is threaded by an ordered poloidal magnetic field corresponding to an initial midplane plasma-$\beta$ of $10^3$. This is considered rather strong in the context of disk dynamics. It is natural to ask whether it is feasible for disks to have such a field in the context of star formation. The strength of the magnetic field in star formation is usually measured by the dimensionless mass-to-flux ratio, $\lambda$, defined as the ratio of the magnetic flux threading a given region to the mass of the same region, normalized to the critical value $(2\pi G^{1/2})^{-1}$ \citep{1978PASJ...30..671N,1997ApJ...475..251S}. If $\lambda < 1$, the magnetic field would be able to support the region against gravitational collapse by itself. Only the so-called `magnetically supercritical' regions with $\lambda > 1$ can collapse and form stars. Zeeman observations have shown that the dense star-forming cores of molecular clouds typically have a dimensionless mass-to-flux ratio of a few after correcting for projection effects ($\lambda_\mathrm{core}\sim 2$; \citealt{2008ApJ...680..457T}). Therefore, the magnetic flux threading a typical low-mass star forming core of 1~$\msun$ would be roughly $\Psi_\mathrm{core}\sim 1.6\times 10^{30}~\mathrm{G~cm}^2$. This is the maximum amount of magnetic flux that is in principle available to the disk. It is much larger than the magnetic flux threading the disk in our reference simulation, which is $\Psi_\mathrm{disk}=1.6\times 10^{28}~\mathrm{G~cm}^2$ (using the fiducial density scaling that yields a relatively massive disk of 0.1~M$_\odot$; it would be smaller by a factor of $\sqrt{10}$ for a 0.01~$\msun$ disk). Therefore, there is no lack of magnetic flux that can in principle be dragged into the disk. Indeed, the problem is the opposite. It is well-established that if a large fraction of the core magnetic flux is deposited in the vicinity of the forming protostar, it would prevent the formation of a rotationally supported disk completely, leading to the so-called `magnetic braking catastrophe' (see \citealt{2014prpl.conf..173L} for a review). In any case, a tiny fraction (one percent or less) of the magnetic flux from the parental core is enough to provide the degree of disk magnetization adopted in our simulation, although how this fraction is determined physically is an outstanding question in the field of disk formation and evolution.

If a poloidal magnetic field corresponding to $\beta=10^3$ is indeed present in the disk, it would drive a rather active accretion. In the high-resolution reference model, the disk accretion rate is of order $10^{-6}$~$\msunperyr$ for a 0.1~$\msun$ disk (see Fig.~\ref{fig:mdot} and \ref{fig:mdot_tavg}); it would be reduced by a factor of 10 for a 0.01~$\msun$ disk. Such rates are too large for the relatively evolved T Tauri (or Class II) disks, but are typical of younger disks around Class 0 and I protostars (e.g., \citealt{2017ApJ...834..178Y}). One may expect such actively accreting protostellar disks to be highly turbulent, if turbulence induced by the MRI (or some other means) is responsible for the accretion. This is indeed the case for our ideal MHD simulation (model ad-els$\infty$) and, to a lesser extent, the most strongly coupled model with ambipolar diffusion (model ad-els1.25). However, strong turbulence is not required for active accretion. This is where the ambipolar diffusivity (and other magnetic diffusivities) comes into play. It is well known that ambipolar diffusion tends to weaken, or even suppress, the MRI (e.g., \citealt{1994ApJ...421..163B,2011ApJ...736..144B,2013ApJ...769...76B,2013ApJ...764...66S}; see \citealt{2014prpl.conf..411T} for a review). It is consistent with the trend that we find in our parameter survey (Section~\ref{sec:param} and Fig.~\ref{fig:5x4}). What is perhaps less obvious at first sight is that the damping of the (MRI-induced) turbulence by ambipolar diffusion (or other means, such as Ohmic dissipation; see, e.g., \citetalias{2017MNRAS.468.3850S}) does not necessarily lead to a shutdown of the disk accretion. Rather, accretion proceeds through another mode: laminar disk accretion driven by a laminar disk wind (\citetalias{2018MNRAS.477.1239S}; \citealt{2018A&A...617A.117R}). In hindsight, it is perhaps not too surprising that ordered, laminar disk winds can drive fast disk accretion. Indeed, early semi-analytic non-ideal MHD models of coupled disk-wind systems tend to run into the opposite problem: if a magnetocentrifugal disk wind is to explain the fast jets observed in young stars, it would require a strong disk magnetization corresponding to $\beta\sim 1$ and would drive the disk to accrete at an uncomfortably high level with accretion velocities comparable to the disk sound speed (see, e.g., \citealt{1993ApJ...410..218W}, their figure~1; \citealt{1995A&A...295..807F,1996ApJ...465..855L}). The disk accretion speed can be reduced by weakening the magnetic field strength (i.e., increasing  plasma-$\beta$), but for a weakly magnetized disk with $\beta\gg 1$ the wind is no longer driven magnetocentrifugally (as beads on a rigid wire; \citealt{1982MNRAS.199..883B}). Instead, the wind is driven by the pressure gradient from a predominantly toroidal magnetic field (see, e.g., \citealt{2016ApJ...818..152B}, the discussion in Section~\ref{sec:global}, and Fig.~\ref{fig:panels}c and~\ref{fig:alongBfield}d for illustration).\footnote{Note, however, that the two views of wind acceleration (centrifugal vs. magnetic) are mathematically equivalent, as discussed in detail in section 4.1 of \citet{1996astro.ph..2022S}.} Although kink instabilities are widely expected in such toroidally dominated magnetic field configurations, there is no clear evidence for their presence in our 3D simulations. The net result is that in the presence of a relatively large ambipolar diffusivity, both the wind and the actively accreting disk are largely laminar. This is very different from the dynamical state of the traditional disk driven turbulent by MRI, as already recognized in the literature (see \citealt{2015arXiv150906382A} for a review; his section~5.4.4). This difference is expected to have implications for the dust dynamics and size evolution, which are closely tied to the gas dynamics.

\subsection{Actively accreting, highly structured, laminar protostellar disks}

An actively accreting laminar disk does not have to be featureless, however. For example, \citet{2016A&A...596A..81P} have shown that the dust in the disk of HL Tau must be settled to a vertical extent of only 1~au at radii as far out as 100~au, implying that the turbulent $\alpha$ parameter must be on the order of $10^{-4}$. As HL Tau is the archetype for accreting protoplanetary disks with axisymmetric substructure, it is now clear that turbulence in circumstellar disks is not a prerequisite for active accretion.

We have shown through both 2D (axisymmetric, \citetalias{2018MNRAS.477.1239S}) and 3D simulations (this paper) that two important types of disk features develop naturally in the wind-launching AD-dominated non-ideal MHD disk: (1) mass accretion concentrated to a thin layer near the disk midplane (see, e.g., Fig.~\ref{fig:panels}d), and (2) the formation of rings and gaps. We attribute the former mostly to ambipolar diffusion, which weakens the MRI and thus allows a laminar flow to develop in the first place. Furthermore, it helps steepen the vertical gradient of the toroidal magnetic field $B_\phi$ (via the \citet{1994ApJ...427L..91B} mechanism) and, thus, the radial component of the current density, which is responsible for driving the disk accretion through a magnetic torque. The latter follows naturally from the former, which enables the sharp pinching of the poloidal field lines in the radial direction. This, in turn, leads to magnetic reconnection at the heart of ring and gap formation, through the redistribution of the poloidal magnetic flux relative to the disk material. The sharp pinching of magnetic field lines is also present in the azimuthal direction. This could in principle lead to the reconnection of the sharply reversed toroidal magnetic field, the driver of the fast midplane accretion. The reconnection of toroidal magnetic field lines at the disk midplane should be captured in these 3D simulations, however, we have shown that the formation of both types of AD-induced substructures -- the fast midplane accretion layer and axisymmetric rings and gaps -- are still possible in three dimensions.

There are several implications for highly structured yet laminar accreting disks. First, the lack of strong turbulence in the disk, especially above and below the midplane accretion layer, makes it easier for the dust grains to settle towards the midplane. This is expected to be particularly significant for protostellar disks that accrete at a high rate; such disks were expected to be more turbulent in the traditional picture of MRI-driven accretion. Second, the dust grains that have settled to the midplane can in principle be moved quickly inward by the fast gas accretion in the thin layer near the midplane through dust-gas drag. This is in addition to the natural tendency for dust grains to drift inward relative to the gas due to their faster orbital speeds and the resultant `head-winds' in the $-\phi$ direction. The dust accretion is expected to be faster through the low-surface density gaps than in the rings both because the gas accretion speed there tends to be higher, and because the radial dust-gas drift speed increases with the Stokes number for grains in the size range most relevant to radio observations (of order 1 cm or less). The Stokes number is inversely proportional to the gas surface density and thus larger (and closer to unity) in the gaps (e.g., \citealt{2015arXiv150906382A}, their section~7.1). Third, and perhaps most importantly, such fast radially migrating dust grains are not expected to be lost from the laminar outer part of the disk quickly. They should be parked in the overdense rings where the gas accretion speed is lower and the gas pressure is at a local maximum in the radial direction; pressure maxima are well-known dust traps (\citealt{1972fpp..conf..211W}; see also section~8.2 of \citealt{2015arXiv150906382A}).

An important point is that the rings (and their associated clumps) are produced through our mechanism early in the process of star formation, when the bulk of the gas and dust is processed through the disk. Indeed, the rings (and gaps) are an inevitable byproduct of the localized midplane accretion driven by the disk wind in an AD-dominated disk threaded by a global poloidal field. If the grains can grow and settle quickly in the disk during the early (Class 0 and I) phases, it may be possible to start trapping the grains much earlier than the protoplanetary (Class II or later) phase, when much more dust is available. The notion of a highly structured yet laminar protostellar disk opens up the possibility that planetesimals (and perhaps planets) could form early in disks during the Class 0 and I phases of star formation. 

In addition, the thermal structure of the disk is expected to be affected by these novel disk features, especially in the thin midplane accretion layer, where the release of the gravitational binding energy is concentrated and the heating by ambipolar diffusion and magnetic reconnection is the most intense. The thermal structure could thus have important implications for the disk chemistry (including snow lines) that should be explored in the future. Furthermore, the concentration of magnetic diffusion and reconnection in a thin midplane accretion layer has implications on how the poloidal magnetic flux is transported throughout the disk, which is an important unsolved problem in disk formation and evolution (e.g., \citealt{2017ApJ...836...46B}).

\section{Summary}\label{sec:conc}

We have carried out a set of 3D non-ideal MHD simulations of magnetically coupled disk-wind systems with different values of ambipolar diffusivity and magnetic field strength. Our main conclusions are as follows.

\begin{enumerate}

\item We illustrated the formation of prominent rings and gaps in 3D in the presence of a moderate level of ambipolar diffusion (with an Elsasser number of order unity at 10~au) and a relatively strong magnetic field (corresponding to $\beta=10^3$). These disk substructures are formed through the same mechanism identified previously in the axisymmetric case: the redistribution of the poloidal magnetic flux relative to the disk material via the reconnection of highly pinched poloidal magnetic field lines in the radial direction. The redistribution is shown clearly in the anti-correlation of the disk surface density and the vertical magnetic field strength at the midplane, with rings of enhanced surface density less strongly magnetized compared to the gaps. 

\item The rings and gaps that develop from the axisymmetric initial conditions adopted in the 3D reference simulation remain nearly axisymmetric at early times. Significant azimuthal variations do develop at later times, especially at relatively large radii where the magnetic field is better coupled to the bulk disk material. The variations may be caused by the development of azimuthal modes of the magnetorotational instability (or other instabilities)  and/or reconnection of a highly pinched magnetic field in the toroidal direction. They do not grow to such an extent as to disrupt the rings and gaps completely, however. The largely coherent rings of enhanced surface density persist to the end of the reference simulation, which lasts for 1500 orbits at the inner disk edge. 

\item We demonstrated through a set of simulations that there are two modes for the coupled disk-wind system in 3D depending on the level of ambipolar diffusivity, as in the axisymmetric (2D) case. In magnetically well-coupled systems with zero or low diffusivities, both the disk and the wind are become chaotic by the development of the so-called `avalanche accretion streams,' a variant of the magnetorotational instability. Highly variable, clumpy substructures develop in the distributions of both the poloidal magnetic field and surface density, particularly in the form of tightly-wound spiral arms. As the ambipolar diffusivity increases, both the disk and the wind become more laminar, with the disk accretion concentrating in a midplane layer and the disk substructures becoming more coherent with more axisymmetric rings and gaps. The disk substructures also depend on the initial magnetic field strength, becoming less prominent in a more weakly magnetized system.

\item The wind-driven laminar mode of disk accretion in the presence of relatively strong ambipolar diffusion has implications for dust dynamics and grain size evolution. The lack of turbulence would allow dust grains to settle towards the midplane more easily and the fast gas accretion in the midplane current layer may advect the settled grains inward quickly. The radially migrating grains may be trapped, however, in the dense rings that develop naturally in such systems. It will be interesting to quantify how the grains settle, migrate, and grow in a laminar yet highly structured disk, especially during the early phases of star formation.

\end{enumerate}

\section*{Acknowledgements}
This work is supported in part by National Science Foundation (NSF) grants AST-1313083 and AST-1716259, National Aeronautics and Space Administration (NASA) grants 80NSSC18K1095 and NNX14AB38G, and Grants-in-Aid for Scientific Research from the MEXT of Japan, 17H01105.

\bibliography{main}{}

\begin{thebibliography}{}
\makeatletter
\relax
\def\mn@urlcharsother{\let\do\@makeother \do\$\do\&\do\#\do\^\do\_\do\%\do\~}
\def\mn@doi{\begingroup\mn@urlcharsother \@ifnextchar [ {\mn@doi@}
  {\mn@doi@[]}}
\def\mn@doi@[#1]#2{\def\@tempa{#1}\ifx\@tempa\@empty \href
  {http://dx.doi.org/#2} {doi:#2}\else \href {http://dx.doi.org/#2} {#1}\fi
  \endgroup}
\def\mn@eprint#1#2{\mn@eprint@#1:#2::\@nil}
\def\mn@eprint@arXiv#1{\href {http://arxiv.org/abs/#1} {{\tt arXiv:#1}}}
\def\mn@eprint@dblp#1{\href {http://dblp.uni-trier.de/rec/bibtex/#1.xml}
  {dblp:#1}}
\def\mn@eprint@#1:#2:#3:#4\@nil{\def\@tempa {#1}\def\@tempb {#2}\def\@tempc
  {#3}\ifx \@tempc \@empty \let \@tempc \@tempb \let \@tempb \@tempa \fi \ifx
  \@tempb \@empty \def\@tempb {arXiv}\fi \@ifundefined
  {mn@eprint@\@tempb}{\@tempb:\@tempc}{\expandafter \expandafter \csname
  mn@eprint@\@tempb\endcsname \expandafter{\@tempc}}}

\bibitem[\protect\citeauthoryear{{ALMA Partnership} et~al.,}{{ALMA Partnership}
  et~al.}{2015}]{2015ApJ...808L...3A}
{ALMA Partnership} et~al., 2015, \mn@doi [\apjl] {10.1088/2041-8205/808/1/L3},
  \href {http://adsabs.harvard.edu/abs/2015ApJ...808L...3A} {808, L3}

\bibitem[\protect\citeauthoryear{{Anderson}, {Li}, {Krasnopolsky}  \&
  {Blandford}}{{Anderson} et~al.}{2006}]{2006ApJ...653L..33A}
{Anderson} J.~M.,  {Li} Z.-Y.,  {Krasnopolsky} R.,   {Blandford} R.~D.,  2006,
  \mn@doi [\apjl] {10.1086/510307}, \href
  {http://adsabs.harvard.edu/abs/2006ApJ...653L..33A} {653, L33}

\bibitem[\protect\citeauthoryear{{Andrews} et~al.,}{{Andrews}
  et~al.}{2016}]{2016ApJ...820L..40A}
{Andrews} S.~M.,  et~al., 2016, \mn@doi [\apjl] {10.3847/2041-8205/820/2/L40},
  \href {http://adsabs.harvard.edu/abs/2016ApJ...820L..40A} {820, L40}

\bibitem[\protect\citeauthoryear{{Armitage}}{{Armitage}}{2015}]{2015arXiv150906382A}
{Armitage} P.~J.,  2015, preprint, \href
  {http://adsabs.harvard.edu/abs/2015arXiv150906382A} {} (\mn@eprint {arXiv}
  {1509.06382})

\bibitem[\protect\citeauthoryear{{Bae}, {Zhu}  \& {Hartmann}}{{Bae}
  et~al.}{2017}]{2017ApJ...850..201B}
{Bae} J.,  {Zhu} Z.,   {Hartmann} L.,  2017, \mn@doi [\apj]
  {10.3847/1538-4357/aa9705}, \href
  {http://adsabs.harvard.edu/abs/2017ApJ...850..201B} {850, 201}

\bibitem[\protect\citeauthoryear{{Bai}}{{Bai}}{2015}]{2015ApJ...798...84B}
{Bai} X.-N.,  2015, \mn@doi [\apj] {10.1088/0004-637X/798/2/84}, \href
  {http://adsabs.harvard.edu/abs/2015ApJ...798...84B} {798, 84}

\bibitem[\protect\citeauthoryear{{Bai}}{{Bai}}{2017}]{2017ApJ...845...75B}
{Bai} X.-N.,  2017, \mn@doi [\apj] {10.3847/1538-4357/aa7dda}, \href
  {http://adsabs.harvard.edu/abs/2017ApJ...845...75B} {845, 75}

\bibitem[\protect\citeauthoryear{{Bai} \& {Goodman}}{{Bai} \&
  {Goodman}}{2009}]{2009ApJ...701..737B}
{Bai} X.-N.,  {Goodman} J.,  2009, \mn@doi [\apj]
  {10.1088/0004-637X/701/1/737}, \href
  {http://adsabs.harvard.edu/abs/2009ApJ...701..737B} {701, 737}

\bibitem[\protect\citeauthoryear{{Bai} \& {Stone}}{{Bai} \&
  {Stone}}{2011}]{2011ApJ...736..144B}
{Bai} X.-N.,  {Stone} J.~M.,  2011, \mn@doi [\apj]
  {10.1088/0004-637X/736/2/144}, \href
  {http://adsabs.harvard.edu/abs/2011ApJ...736..144B} {736, 144}

\bibitem[\protect\citeauthoryear{{Bai} \& {Stone}}{{Bai} \&
  {Stone}}{2013}]{2013ApJ...769...76B}
{Bai} X.-N.,  {Stone} J.~M.,  2013, \mn@doi [\apj]
  {10.1088/0004-637X/769/1/76}, \href
  {http://adsabs.harvard.edu/abs/2013ApJ...769...76B} {769, 76}

\bibitem[\protect\citeauthoryear{{Bai} \& {Stone}}{{Bai} \&
  {Stone}}{2014}]{2014ApJ...796...31B}
{Bai} X.-N.,  {Stone} J.~M.,  2014, \mn@doi [\apj]
  {10.1088/0004-637X/796/1/31}, \href
  {http://adsabs.harvard.edu/abs/2014ApJ...796...31B} {796, 31}

\bibitem[\protect\citeauthoryear{{Bai} \& {Stone}}{{Bai} \&
  {Stone}}{2017}]{2017ApJ...836...46B}
{Bai} X.-N.,  {Stone} J.~M.,  2017, \mn@doi [\apj]
  {10.3847/1538-4357/836/1/46}, \href
  {http://adsabs.harvard.edu/abs/2017ApJ...836...46B} {836, 46}

\bibitem[\protect\citeauthoryear{{Bai}, {Ye}, {Goodman}  \& {Yuan}}{{Bai}
  et~al.}{2016}]{2016ApJ...818..152B}
{Bai} X.-N.,  {Ye} J.,  {Goodman} J.,   {Yuan} F.,  2016, \mn@doi [\apj]
  {10.3847/0004-637X/818/2/152}, \href
  {http://adsabs.harvard.edu/abs/2016ApJ...818..152B} {818, 152}

\bibitem[\protect\citeauthoryear{{Balbus} \& {Hawley}}{{Balbus} \&
  {Hawley}}{1992}]{1992ApJ...400..610B}
{Balbus} S.~A.,  {Hawley} J.~F.,  1992, \mn@doi [\apj] {10.1086/172022}, \href
  {http://adsabs.harvard.edu/abs/1992ApJ...400..610B} {400, 610}

\bibitem[\protect\citeauthoryear{{Banzatti}, {Pascucci}, {Edwards}, {Fang},
  {Gorti}  \& {Flock}}{{Banzatti} et~al.}{2018}]{2018arXiv181106544B}
{Banzatti} A.,  {Pascucci} I.,  {Edwards} S.,  {Fang} M.,  {Gorti} U.,
  {Flock} M.,  2018, preprint, \href
  {http://adsabs.harvard.edu/abs/2018arXiv181106544B} {} (\mn@eprint {arXiv}
  {1811.06544})

\bibitem[\protect\citeauthoryear{{B{\'e}thune}, {Lesur}  \&
  {Ferreira}}{{B{\'e}thune} et~al.}{2016}]{2016A&A...589A..87B}
{B{\'e}thune} W.,  {Lesur} G.,   {Ferreira} J.,  2016, \mn@doi [\aap]
  {10.1051/0004-6361/201527874}, \href
  {http://adsabs.harvard.edu/abs/2016A%26A...589A..87B} {589, A87}

\bibitem[\protect\citeauthoryear{{B{\'e}thune}, {Lesur}  \&
  {Ferreira}}{{B{\'e}thune} et~al.}{2017}]{2017A&A...600A..75B}
{B{\'e}thune} W.,  {Lesur} G.,   {Ferreira} J.,  2017, \mn@doi [\aap]
  {10.1051/0004-6361/201630056}, \href
  {http://adsabs.harvard.edu/abs/2017A%26A...600A..75B} {600, A75}

\bibitem[\protect\citeauthoryear{{Bjerkeli}, {van der Wiel}, {Harsono},
  {Ramsey}  \& {J{\o}rgensen}}{{Bjerkeli} et~al.}{2016}]{2016Natur.540..406B}
{Bjerkeli} P.,  {van der Wiel} M.~H.~D.,  {Harsono} D.,  {Ramsey} J.~P.,
  {J{\o}rgensen} J.~K.,  2016, \mn@doi [\nat] {10.1038/nature20600}, \href
  {http://adsabs.harvard.edu/abs/2016Natur.540..406B} {540, 406}

\bibitem[\protect\citeauthoryear{{Blaes} \& {Balbus}}{{Blaes} \&
  {Balbus}}{1994}]{1994ApJ...421..163B}
{Blaes} O.~M.,  {Balbus} S.~A.,  1994, \mn@doi [\apj] {10.1086/173634}, \href
  {http://adsabs.harvard.edu/abs/1994ApJ...421..163B} {421, 163}

\bibitem[\protect\citeauthoryear{{Blandford} \& {Payne}}{{Blandford} \&
  {Payne}}{1982}]{1982MNRAS.199..883B}
{Blandford} R.~D.,  {Payne} D.~G.,  1982, \mn@doi [\mnras]
  {10.1093/mnras/199.4.883}, \href
  {http://adsabs.harvard.edu/abs/1982MNRAS.199..883B} {199, 883}

\bibitem[\protect\citeauthoryear{{Brandenburg} \& {Zweibel}}{{Brandenburg} \&
  {Zweibel}}{1994}]{1994ApJ...427L..91B}
{Brandenburg} A.,  {Zweibel} E.~G.,  1994, \mn@doi [\apjl] {10.1086/187372},
  \href {http://adsabs.harvard.edu/abs/1994ApJ...427L..91B} {427, L91}

\bibitem[\protect\citeauthoryear{{Cieza} et~al.,}{{Cieza}
  et~al.}{2016}]{2016Natur.535..258C}
{Cieza} L.~A.,  et~al., 2016, \mn@doi [\nat] {10.1038/nature18612}, \href
  {http://adsabs.harvard.edu/abs/2016Natur.535..258C} {535, 258}

\bibitem[\protect\citeauthoryear{{Dipierro} et~al.,}{{Dipierro}
  et~al.}{2018}]{2018MNRAS.475.5296D}
{Dipierro} G.,  et~al., 2018, \mn@doi [\mnras] {10.1093/mnras/sty181}, \href
  {http://adsabs.harvard.edu/abs/2018MNRAS.475.5296D} {475, 5296}

\bibitem[\protect\citeauthoryear{{Dong}, {Zhu}  \& {Whitney}}{{Dong}
  et~al.}{2015}]{2015ApJ...809...93D}
{Dong} R.,  {Zhu} Z.,   {Whitney} B.,  2015, \mn@doi [\apj]
  {10.1088/0004-637X/809/1/93}, \href
  {http://adsabs.harvard.edu/abs/2015ApJ...809...93D} {809, 93}

\bibitem[\protect\citeauthoryear{{Dong}, {Li}, {Chiang}  \& {Li}}{{Dong}
  et~al.}{2017}]{2017ApJ...843..127D}
{Dong} R.,  {Li} S.,  {Chiang} E.,   {Li} H.,  2017, \mn@doi [\apj]
  {10.3847/1538-4357/aa72f2}, \href
  {http://adsabs.harvard.edu/abs/2017ApJ...843..127D} {843, 127}

\bibitem[\protect\citeauthoryear{{Dzyurkevich}, {Flock}, {Turner}, {Klahr}  \&
  {Henning}}{{Dzyurkevich} et~al.}{2010}]{2010A&A...515A..70D}
{Dzyurkevich} N.,  {Flock} M.,  {Turner} N.~J.,  {Klahr} H.,   {Henning} T.,
  2010, \mn@doi [\aap] {10.1051/0004-6361/200912834}, \href
  {http://adsabs.harvard.edu/abs/2010A%26A...515A..70D} {515, A70}

\bibitem[\protect\citeauthoryear{{Fedele} et~al.,}{{Fedele}
  et~al.}{2017}]{2017A&A...600A..72F}
{Fedele} D.,  et~al., 2017, \mn@doi [\aap] {10.1051/0004-6361/201629860}, \href
  {http://adsabs.harvard.edu/abs/2017A%26A...600A..72F} {600, A72}

\bibitem[\protect\citeauthoryear{{Fedele} et~al.,}{{Fedele}
  et~al.}{2018}]{2018A&A...610A..24F}
{Fedele} D.,  et~al., 2018, \mn@doi [\aap] {10.1051/0004-6361/201731978}, \href
  {http://adsabs.harvard.edu/abs/2018A%26A...610A..24F} {610, A24}

\bibitem[\protect\citeauthoryear{{Ferreira} \& {Pelletier}}{{Ferreira} \&
  {Pelletier}}{1995}]{1995A&A...295..807F}
{Ferreira} J.,  {Pelletier} G.,  1995, \aap, \href
  {http://adsabs.harvard.edu/abs/1995A%26A...295..807F} {295, 807}

\bibitem[\protect\citeauthoryear{{Flock}, {Dzyurkevich}, {Klahr}, {Turner}  \&
  {Henning}}{{Flock} et~al.}{2011}]{2011ApJ...735..122F}
{Flock} M.,  {Dzyurkevich} N.,  {Klahr} H.,  {Turner} N.~J.,   {Henning} T.,
  2011, \mn@doi [\apj] {10.1088/0004-637X/735/2/122}, \href
  {http://adsabs.harvard.edu/abs/2011ApJ...735..122F} {735, 122}

\bibitem[\protect\citeauthoryear{{Flock}, {Ruge}, {Dzyurkevich}, {Henning},
  {Klahr}  \& {Wolf}}{{Flock} et~al.}{2015}]{2015A&A...574A..68F}
{Flock} M.,  {Ruge} J.~P.,  {Dzyurkevich} N.,  {Henning} T.,  {Klahr} H.,
  {Wolf} S.,  2015, \mn@doi [\aap] {10.1051/0004-6361/201424693}, \href
  {http://adsabs.harvard.edu/abs/2015A%26A...574A..68F} {574, A68}

\bibitem[\protect\citeauthoryear{{Glassgold}, {Lizano}  \& {Galli}}{{Glassgold}
  et~al.}{2017}]{2017MNRAS.472.2447G}
{Glassgold} A.~E.,  {Lizano} S.,   {Galli} D.,  2017, \mn@doi [\mnras]
  {10.1093/mnras/stx2145}, \href
  {http://adsabs.harvard.edu/abs/2017MNRAS.472.2447G} {472, 2447}

\bibitem[\protect\citeauthoryear{{Goodman} \& {Xu}}{{Goodman} \&
  {Xu}}{1994}]{1994ApJ...432..213G}
{Goodman} J.,  {Xu} G.,  1994, \mn@doi [\apj] {10.1086/174562}, \href
  {http://adsabs.harvard.edu/abs/1994ApJ...432..213G} {432, 213}

\bibitem[\protect\citeauthoryear{{Hawley}}{{Hawley}}{2000}]{2000ApJ...528..462H}
{Hawley} J.~F.,  2000, \mn@doi [\apj] {10.1086/308180}, \href
  {http://adsabs.harvard.edu/abs/2000ApJ...528..462H} {528, 462}

\bibitem[\protect\citeauthoryear{{Hawley}, {Gammie}  \& {Balbus}}{{Hawley}
  et~al.}{1995}]{1995ApJ...440..742H}
{Hawley} J.~F.,  {Gammie} C.~F.,   {Balbus} S.~A.,  1995, \mn@doi [\apj]
  {10.1086/175311}, \href {http://adsabs.harvard.edu/abs/1995ApJ...440..742H}
  {440, 742}

\bibitem[\protect\citeauthoryear{{Isella} et~al.,}{{Isella}
  et~al.}{2016}]{2016PhRvL.117y1101I}
{Isella} A.,  et~al., 2016, \mn@doi [Physical Review Letters]
  {10.1103/PhysRevLett.117.251101}, \href
  {http://adsabs.harvard.edu/abs/2016PhRvL.117y1101I} {117, 251101}

\bibitem[\protect\citeauthoryear{{Johansen}, {Youdin}  \& {Klahr}}{{Johansen}
  et~al.}{2009}]{2009ApJ...697.1269J}
{Johansen} A.,  {Youdin} A.,   {Klahr} H.,  2009, \mn@doi [\apj]
  {10.1088/0004-637X/697/2/1269}, \href
  {http://adsabs.harvard.edu/abs/2009ApJ...697.1269J} {697, 1269}

\bibitem[\protect\citeauthoryear{{Krapp}, {Gressel}, {Ben{\'{\i}}tez-Llambay},
  {Downes}, {Mohandas}  \& {Pessah}}{{Krapp}
  et~al.}{2018}]{2018ApJ...865..105K}
{Krapp} L.,  {Gressel} O.,  {Ben{\'{\i}}tez-Llambay} P.,  {Downes} T.~P.,
  {Mohandas} G.,   {Pessah} M.~E.,  2018, \mn@doi [\apj]
  {10.3847/1538-4357/aadcf0}, \href
  {http://adsabs.harvard.edu/abs/2018ApJ...865..105K} {865, 105}

\bibitem[\protect\citeauthoryear{{Krasnopolsky}, {Li}  \&
  {Shang}}{{Krasnopolsky} et~al.}{2010}]{2010ApJ...716.1541K}
{Krasnopolsky} R.,  {Li} Z.-Y.,   {Shang} H.,  2010, \mn@doi [\apj]
  {10.1088/0004-637X/716/2/1541}, \href
  {http://adsabs.harvard.edu/abs/2010ApJ...716.1541K} {716, 1541}

\bibitem[\protect\citeauthoryear{{Kunz} \& {Lesur}}{{Kunz} \&
  {Lesur}}{2013}]{2013MNRAS.434.2295K}
{Kunz} M.~W.,  {Lesur} G.,  2013, \mn@doi [\mnras] {10.1093/mnras/stt1171},
  \href {http://adsabs.harvard.edu/abs/2013MNRAS.434.2295K} {434, 2295}

\bibitem[\protect\citeauthoryear{{Li}}{{Li}}{1996}]{1996ApJ...465..855L}
{Li} Z.-Y.,  1996, \mn@doi [\apj] {10.1086/177469}, \href
  {http://adsabs.harvard.edu/abs/1996ApJ...465..855L} {465, 855}

\bibitem[\protect\citeauthoryear{{Li}, {Banerjee}, {Pudritz}, {J{\o}rgensen},
  {Shang}, {Krasnopolsky}  \& {Maury}}{{Li} et~al.}{2014}]{2014prpl.conf..173L}
{Li} Z.-Y.,  {Banerjee} R.,  {Pudritz} R.~E.,  {J{\o}rgensen} J.~K.,  {Shang}
  H.,  {Krasnopolsky} R.,   {Maury} A.,  2014, \mn@doi [Protostars and Planets
  VI] {10.2458/azu_uapress_9780816531240-ch008}, \href
  {http://adsabs.harvard.edu/abs/2014prpl.conf..173L} {pp 173--194}

\bibitem[\protect\citeauthoryear{{Lovelace}, {Li}, {Colgate}  \&
  {Nelson}}{{Lovelace} et~al.}{1999}]{1999ApJ...513..805L}
{Lovelace} R.~V.~E.,  {Li} H.,  {Colgate} S.~A.,   {Nelson} A.~F.,  1999,
  \mn@doi [\apj] {10.1086/306900}, \href
  {http://adsabs.harvard.edu/abs/1999ApJ...513..805L} {513, 805}

\bibitem[\protect\citeauthoryear{{Moll}}{{Moll}}{2012}]{2012A&A...548A..76M}
{Moll} R.,  2012, \mn@doi [\aap] {10.1051/0004-6361/201118249}, \href
  {http://adsabs.harvard.edu/abs/2012A%26A...548A..76M} {548, A76}

\bibitem[\protect\citeauthoryear{{Nakano} \& {Nakamura}}{{Nakano} \&
  {Nakamura}}{1978}]{1978PASJ...30..671N}
{Nakano} T.,  {Nakamura} T.,  1978, \pasj, \href
  {http://adsabs.harvard.edu/abs/1978PASJ...30..671N} {30, 671}

\bibitem[\protect\citeauthoryear{{Nomura} et~al.,}{{Nomura}
  et~al.}{2016}]{2016ApJ...819L...7N}
{Nomura} H.,  et~al., 2016, \mn@doi [\apjl] {10.3847/2041-8205/819/1/L7}, \href
  {http://adsabs.harvard.edu/abs/2016ApJ...819L...7N} {819, L7}

\bibitem[\protect\citeauthoryear{{Okuzumi}, {Momose}, {Sirono}, {Kobayashi}  \&
  {Tanaka}}{{Okuzumi} et~al.}{2016}]{2016ApJ...821...82O}
{Okuzumi} S.,  {Momose} M.,  {Sirono} S.-i.,  {Kobayashi} H.,   {Tanaka} H.,
  2016, \mn@doi [\apj] {10.3847/0004-637X/821/2/82}, \href
  {http://adsabs.harvard.edu/abs/2016ApJ...821...82O} {821, 82}

\bibitem[\protect\citeauthoryear{{Parker}}{{Parker}}{1957}]{1957JGR....62..509P}
{Parker} E.~N.,  1957, \mn@doi [\jgr] {10.1029/JZ062i004p00509}, \href
  {http://adsabs.harvard.edu/abs/1957JGR....62..509P} {62, 509}

\bibitem[\protect\citeauthoryear{{Perez-Becker} \& {Chiang}}{{Perez-Becker} \&
  {Chiang}}{2011}]{2011ApJ...735....8P}
{Perez-Becker} D.,  {Chiang} E.,  2011, \mn@doi [\apj]
  {10.1088/0004-637X/735/1/8}, \href
  {http://adsabs.harvard.edu/abs/2011ApJ...735....8P} {735, 8}

\bibitem[\protect\citeauthoryear{{P{\'e}rez} et~al.,}{{P{\'e}rez}
  et~al.}{2016}]{2016Sci...353.1519P}
{P{\'e}rez} L.~M.,  et~al., 2016, \mn@doi [Science] {10.1126/science.aaf8296},
  \href {http://adsabs.harvard.edu/abs/2016Sci...353.1519P} {353, 1519}

\bibitem[\protect\citeauthoryear{{Pinilla}, {Flock}, {Ovelar}  \&
  {Birnstiel}}{{Pinilla} et~al.}{2016}]{2016A&A...596A..81P}
{Pinilla} P.,  {Flock} M.,  {Ovelar} M.~d.~J.,   {Birnstiel} T.,  2016, \mn@doi
  [\aap] {10.1051/0004-6361/201628441}, \href
  {http://adsabs.harvard.edu/abs/2016A%26A...596A..81P} {596, A81}

\bibitem[\protect\citeauthoryear{{Riols} \& {Lesur}}{{Riols} \&
  {Lesur}}{2018}]{2018A&A...617A.117R}
{Riols} A.,  {Lesur} G.,  2018, \mn@doi [\aap] {10.1051/0004-6361/201833212},
  \href {http://adsabs.harvard.edu/abs/2018A%26A...617A.117R} {617, A117}

\bibitem[\protect\citeauthoryear{{Ruge}, {Flock}, {Wolf}, {Dzyurkevich},
  {Fromang}, {Henning}, {Klahr}  \& {Meheut}}{{Ruge}
  et~al.}{2016}]{2016A&A...590A..17R}
{Ruge} J.~P.,  {Flock} M.,  {Wolf} S.,  {Dzyurkevich} N.,  {Fromang} S.,
  {Henning} T.,  {Klahr} H.,   {Meheut} H.,  2016, \mn@doi [\aap]
  {10.1051/0004-6361/201526616}, \href
  {http://adsabs.harvard.edu/abs/2016A%26A...590A..17R} {590, A17}

\bibitem[\protect\citeauthoryear{{Shakura} \& {Sunyaev}}{{Shakura} \&
  {Sunyaev}}{1973}]{1973A&A....24..337S}
{Shakura} N.~I.,  {Sunyaev} R.~A.,  1973, \aap, \href
  {http://adsabs.harvard.edu/abs/1973A%26A....24..337S} {24, 337}

\bibitem[\protect\citeauthoryear{{Shu}}{{Shu}}{1992}]{1992phas.book.....S}
{Shu} F.~H.,  1992, {Physics of Astrophysics, Vol. II}.
University Science Books

\bibitem[\protect\citeauthoryear{{Shu} \& {Li}}{{Shu} \&
  {Li}}{1997}]{1997ApJ...475..251S}
{Shu} F.~H.,  {Li} Z.-Y.,  1997, \mn@doi [\apj] {10.1086/303521}, \href
  {http://adsabs.harvard.edu/abs/1997ApJ...475..251S} {475, 251}

\bibitem[\protect\citeauthoryear{{Simon} \& {Armitage}}{{Simon} \&
  {Armitage}}{2014}]{2014ApJ...784...15S}
{Simon} J.~B.,  {Armitage} P.~J.,  2014, \mn@doi [\apj]
  {10.1088/0004-637X/784/1/15}, \href
  {http://adsabs.harvard.edu/abs/2014ApJ...784...15S} {784, 15}

\bibitem[\protect\citeauthoryear{{Simon}, {Bai}, {Stone}, {Armitage}  \&
  {Beckwith}}{{Simon} et~al.}{2013a}]{2013ApJ...764...66S}
{Simon} J.~B.,  {Bai} X.-N.,  {Stone} J.~M.,  {Armitage} P.~J.,   {Beckwith}
  K.,  2013a, \mn@doi [\apj] {10.1088/0004-637X/764/1/66}, \href
  {http://adsabs.harvard.edu/abs/2013ApJ...764...66S} {764, 66}

\bibitem[\protect\citeauthoryear{{Simon}, {Bai}, {Armitage}, {Stone}  \&
  {Beckwith}}{{Simon} et~al.}{2013b}]{2013ApJ...775...73S}
{Simon} J.~B.,  {Bai} X.-N.,  {Armitage} P.~J.,  {Stone} J.~M.,   {Beckwith}
  K.,  2013b, \mn@doi [\apj] {10.1088/0004-637X/775/1/73}, \href
  {http://adsabs.harvard.edu/abs/2013ApJ...775...73S} {775, 73}

\bibitem[\protect\citeauthoryear{{Spruit}}{{Spruit}}{1996}]{1996astro.ph..2022S}
{Spruit} H.~C.,  1996, ArXiv Astrophysics e-prints, \href
  {http://adsabs.harvard.edu/abs/1996astro.ph..2022S} {}

\bibitem[\protect\citeauthoryear{{Stepanovs} \& {Fendt}}{{Stepanovs} \&
  {Fendt}}{2014}]{2014ApJ...793...31S}
{Stepanovs} D.,  {Fendt} C.,  2014, \mn@doi [\apj]
  {10.1088/0004-637X/793/1/31}, \href
  {http://adsabs.harvard.edu/abs/2014ApJ...793...31S} {793, 31}

\bibitem[\protect\citeauthoryear{{Suriano}, {Li}, {Krasnopolsky}  \&
  {Shang}}{{Suriano} et~al.}{2017}]{2017MNRAS.468.3850S}
{Suriano} S.~S.,  {Li} Z.-Y.,  {Krasnopolsky} R.,   {Shang} H.,  2017, \mn@doi
  [\mnras] {10.1093/mnras/stx735}, \href
  {http://adsabs.harvard.edu/abs/2017MNRAS.468.3850S} {468, 3850}

\bibitem[\protect\citeauthoryear{{Suriano}, {Li}, {Krasnopolsky}  \&
  {Shang}}{{Suriano} et~al.}{2018}]{2018MNRAS.477.1239S}
{Suriano} S.~S.,  {Li} Z.-Y.,  {Krasnopolsky} R.,   {Shang} H.,  2018, \mn@doi
  [\mnras] {10.1093/mnras/sty717}, \href
  {http://adsabs.harvard.edu/abs/2018MNRAS.477.1239S} {477, 1239}

\bibitem[\protect\citeauthoryear{{Suzuki} \& {Inutsuka}}{{Suzuki} \&
  {Inutsuka}}{2014}]{2014ApJ...784..121S}
{Suzuki} T.~K.,  {Inutsuka} S.-i.,  2014, \mn@doi [\apj]
  {10.1088/0004-637X/784/2/121}, \href
  {http://adsabs.harvard.edu/abs/2014ApJ...784..121S} {784, 121}

\bibitem[\protect\citeauthoryear{{Sweet}}{{Sweet}}{1958}]{1958IAUS....6..123S}
{Sweet} P.~A.,  1958, in {Lehnert} B.,  ed.,  IAU Symposium Vol. 6,
  Electromagnetic Phenomena in Cosmical Physics. p.~123

\bibitem[\protect\citeauthoryear{{Takahashi} \& {Inutsuka}}{{Takahashi} \&
  {Inutsuka}}{2014}]{2014ApJ...794...55T}
{Takahashi} S.~Z.,  {Inutsuka} S.-i.,  2014, \mn@doi [\apj]
  {10.1088/0004-637X/794/1/55}, \href
  {http://adsabs.harvard.edu/abs/2014ApJ...794...55T} {794, 55}

\bibitem[\protect\citeauthoryear{{Takahashi} \& {Inutsuka}}{{Takahashi} \&
  {Inutsuka}}{2016}]{2016AJ....152..184T}
{Takahashi} S.~Z.,  {Inutsuka} S.-i.,  2016, \mn@doi [\aj]
  {10.3847/0004-6256/152/6/184}, \href
  {http://adsabs.harvard.edu/abs/2016AJ....152..184T} {152, 184}

\bibitem[\protect\citeauthoryear{{Takahashi} \& {Muto}}{{Takahashi} \&
  {Muto}}{2018}]{2018ApJ...865..102T}
{Takahashi} S.~Z.,  {Muto} T.,  2018, \mn@doi [\apj]
  {10.3847/1538-4357/aadda0}, \href
  {http://adsabs.harvard.edu/abs/2018ApJ...865..102T} {865, 102}

\bibitem[\protect\citeauthoryear{{Takasao}, {Tomida}, {Iwasaki}  \&
  {Suzuki}}{{Takasao} et~al.}{2018}]{2018ApJ...857....4T}
{Takasao} S.,  {Tomida} K.,  {Iwasaki} K.,   {Suzuki} T.~K.,  2018, \mn@doi
  [\apj] {10.3847/1538-4357/aab5b3}, \href
  {http://adsabs.harvard.edu/abs/2018ApJ...857....4T} {857, 4}

\bibitem[\protect\citeauthoryear{{Troland} \& {Crutcher}}{{Troland} \&
  {Crutcher}}{2008}]{2008ApJ...680..457T}
{Troland} T.~H.,  {Crutcher} R.~M.,  2008, \mn@doi [\apj] {10.1086/587546},
  \href {http://adsabs.harvard.edu/abs/2008ApJ...680..457T} {680, 457}

\bibitem[\protect\citeauthoryear{{Turner}, {Fromang}, {Gammie}, {Klahr},
  {Lesur}, {Wardle}  \& {Bai}}{{Turner} et~al.}{2014}]{2014prpl.conf..411T}
{Turner} N.~J.,  {Fromang} S.,  {Gammie} C.,  {Klahr} H.,  {Lesur} G.,
  {Wardle} M.,   {Bai} X.-N.,  2014, \mn@doi [Protostars and Planets VI]
  {10.2458/azu_uapress_9780816531240-ch018}, \href
  {http://adsabs.harvard.edu/abs/2014prpl.conf..411T} {pp 411--432}

\bibitem[\protect\citeauthoryear{{Umebayashi} \& {Nakano}}{{Umebayashi} \&
  {Nakano}}{1981}]{1981PASJ...33..617U}
{Umebayashi} T.,  {Nakano} T.,  1981, \pasj, \href
  {http://adsabs.harvard.edu/abs/1981PASJ...33..617U} {33, 617}

\bibitem[\protect\citeauthoryear{{Wang}, {Bai}  \& {Goodman}}{{Wang}
  et~al.}{2018}]{2018arXiv181012330W}
{Wang} L.,  {Bai} X.-N.,   {Goodman} J.,  2018, preprint, \href
  {http://adsabs.harvard.edu/abs/2018arXiv181012330W} {} (\mn@eprint {arXiv}
  {1810.12330})

\bibitem[\protect\citeauthoryear{{Wardle} \& {Koenigl}}{{Wardle} \&
  {Koenigl}}{1993}]{1993ApJ...410..218W}
{Wardle} M.,  {Koenigl} A.,  1993, \mn@doi [\apj] {10.1086/172739}, \href
  {http://adsabs.harvard.edu/abs/1993ApJ...410..218W} {410, 218}

\bibitem[\protect\citeauthoryear{{Whipple}}{{Whipple}}{1972}]{1972fpp..conf..211W}
{Whipple} F.~L.,  1972, in {Elvius} A.,  ed., From Plasma to Planet. p.~211

\bibitem[\protect\citeauthoryear{{Yen}, {Koch}, {Takakuwa}, {Krasnopolsky},
  {Ohashi}  \& {Aso}}{{Yen} et~al.}{2017}]{2017ApJ...834..178Y}
{Yen} H.-W.,  {Koch} P.~M.,  {Takakuwa} S.,  {Krasnopolsky} R.,  {Ohashi} N.,
  {Aso} Y.,  2017, \mn@doi [\apj] {10.3847/1538-4357/834/2/178}, \href
  {http://adsabs.harvard.edu/abs/2017ApJ...834..178Y} {834, 178}

\bibitem[\protect\citeauthoryear{{Zanni}, {Ferrari}, {Rosner}, {Bodo}  \&
  {Massaglia}}{{Zanni} et~al.}{2007}]{2007A&A...469..811Z}
{Zanni} C.,  {Ferrari} A.,  {Rosner} R.,  {Bodo} G.,   {Massaglia} S.,  2007,
  \mn@doi [\aap] {10.1051/0004-6361:20066400}, \href
  {http://adsabs.harvard.edu/abs/2007A%26A...469..811Z} {469, 811}

\bibitem[\protect\citeauthoryear{{Zhang}, {Blake}  \& {Bergin}}{{Zhang}
  et~al.}{2015}]{2015ApJ...806L...7Z}
{Zhang} K.,  {Blake} G.~A.,   {Bergin} E.~A.,  2015, \mn@doi [\apjl]
  {10.1088/2041-8205/806/1/L7}, \href
  {http://adsabs.harvard.edu/abs/2015ApJ...806L...7Z} {806, L7}

\bibitem[\protect\citeauthoryear{{Zhang}, {Bergin}, {Blake}, {Cleeves},
  {Hogerheijde}, {Salinas}  \& {Schwarz}}{{Zhang}
  et~al.}{2016}]{2016ApJ...818L..16Z}
{Zhang} K.,  {Bergin} E.~A.,  {Blake} G.~A.,  {Cleeves} L.~I.,  {Hogerheijde}
  M.,  {Salinas} V.,   {Schwarz} K.~R.,  2016, \mn@doi [\apjl]
  {10.3847/2041-8205/818/1/L16}, \href
  {http://adsabs.harvard.edu/abs/2016ApJ...818L..16Z} {818, L16}

\bibitem[\protect\citeauthoryear{{Zhu} \& {Stone}}{{Zhu} \&
  {Stone}}{2018}]{2018ApJ...857...34Z}
{Zhu} Z.,  {Stone} J.~M.,  2018, \mn@doi [\apj] {10.3847/1538-4357/aaafc9},
  \href {http://adsabs.harvard.edu/abs/2018ApJ...857...34Z} {857, 34}

\bibitem[\protect\citeauthoryear{{Zweibel} \& {Yamada}}{{Zweibel} \&
  {Yamada}}{2009}]{2009ARA&A..47..291Z}
{Zweibel} E.~G.,  {Yamada} M.,  2009, \mn@doi [\araa]
  {10.1146/annurev-astro-082708-101726}, \href
  {http://adsabs.harvard.edu/abs/2009ARA%26A..47..291Z} {47, 291}

\bibitem[\protect\citeauthoryear{{van der Plas} et~al.,}{{van der Plas}
  et~al.}{2017}]{2017A&A...597A..32V}
{van der Plas} G.,  et~al., 2017, \mn@doi [\aap] {10.1051/0004-6361/201629523},
  \href {http://adsabs.harvard.edu/abs/2017A%26A...597A..32V} {597, A32}

\makeatother
\end{thebibliography}
\bibliographystyle{mnras}

\bsp	
\label{lastpage}
\end{document}